\documentclass[10pt,twocolumn]{IEEEtran}
\usepackage{graphicx}
\usepackage{amssymb}
\usepackage{float}
\usepackage[cmex10]{amsmath}
\usepackage{cite}
\usepackage{algorithm} 
\usepackage{algpseudocode} 
\usepackage{comment}
\usepackage{url}
\usepackage{xcolor}
\usepackage{hyperref}
\hypersetup{
    colorlinks=true,
    linkcolor=blue,
    citecolor=blue,          
    urlcolor=blue,
    pdftitle={CalibFPA}
}

\usepackage{multirow}

\makeatletter
{}

\makeatother

\usepackage[font=footnotesize,justification=justified,belowskip=2pt,aboveskip=2pt]{caption}

\newcounter{algoCtr}
\stepcounter{algoCtr}

\newcommand{\xb}{\mathbf{x}}
\newcommand{\zb}{\mathbf{z}}

\newcommand{\yb}{\mathbf{y}}

\newcommand{\Db}{\mathbf{D}}

\newcommand{\Ib}{\mathbf{I}}

\newcommand{\db}{\mathbf{d}}

\newcommand{\rb}{\mathbf{r}}

\newcommand{\scaleExperimental}{0.23}
\newcommand{\scaleSimulated}{0.18}
\newcommand{\MinipageExp}{0.155}
\newcommand{\MinipageSim}{0.14}
\newcommand{\verticalSpacingDist}{2px}

\begin{document}
%
\title{CalibFPA: A Focal Plane Array Imaging System based on Online Deep-Learning Calibration}
%
%
%
\author{
Alper~G\"{u}ng\"{o}r$^{*,\dagger}$, M. Umut Bahceci$^*$, Yasin Ergen, Ahmet Sözak, O. Oner Ekiz, \\ Tolga Yelboga, Tolga Çukur 
\thanks{A.~G\"{u}ng\"{o}r, M. U. Bahceci, A. Sozak, T. Yelboga are with ASELSAN, Ankara, Turkiye.} 
\thanks{Y. Ergen and O. O. Ekiz are with Nanodev Scientific, Ankara, Turkiye.} \thanks{O. O. Ekiz is also with dept. of Materials Eng., Ostim Technical University, Ankara, Turkiye.}
\thanks{T. Çukur is with Dept. of Electrical-Electronic Eng., National Magnetic Resonance Research Center (UMRAM), Bilkent University, Ankara, Turkiye.}
\thanks{$^*$These authors contributed equally.}
\thanks{$^\dagger$Corresponding author (email: alpergungor@windowslive.com).}
}

\maketitle

\begin{abstract}
Compressive focal plane arrays (FPA) enable cost-effective high-resolution (HR) imaging by acquisition of several multiplexed measurements on a low-resolution (LR) sensor. Multiplexed encoding of the visual scene is typically performed via electronically controllable spatial light modulators (SLM). An HR image is then reconstructed from the encoded measurements by solving an inverse problem that involves the forward model of the imaging system. To capture system non-idealities such as optical aberrations, a mainstream approach is to conduct an offline calibration scan to measure the system response for a point source at each spatial location on the imaging grid. However, it is challenging to run calibration scans when using structured SLMs as they cannot encode individual grid locations. In this study, we propose a novel compressive FPA system based on online deep-learning calibration of multiplexed LR measurements (CalibFPA). We introduce a piezo-stage that locomotes a pre-printed fixed coded aperture. A deep neural network is then leveraged to correct for the influences of system non-idealities in multiplexed measurements without the need for offline calibration scans. Finally, a deep plug-and-play algorithm is used to reconstruct images from corrected measurements. On simulated and experimental datasets, we demonstrate that CalibFPA outperforms state-of-the-art compressive FPA methods. We also report analyses to validate the design elements in CalibFPA and assess computational complexity.  
\end{abstract}

\IEEEpeerreviewmaketitle


\IEEEpeerreviewmaketitle

\bstctlcite{IEEEexample:BSTcontrol}

\section{Introduction}


Conventional imaging systems leverage high-resolution (HR) sensors with millions of pixels, therefore requiring complex, expensive manufacturing processes \cite{SPI, SPI_hist}. In recent years, multiplexed imaging has come forth as a cost-efficient framework to image HR scenes with low-resolution (LR) sensors for numerous imaging modalities operating across various wavelengths \cite{SPI, duarteSPI, fpa-cs, SPI_hist, Mahalanobis2014, stantchev2020thz, li2022thz}. In this framework, a spatial light modulator (SLM), typically placed in between the imaged object and the LR sensor, performs spatial encoding of the scene \cite{SPI_hist, duarteSPI, paunescu2018compressive, lisens, fpa-cs, cassi, HighAccuracyCodedAperture}. The SLM contains a dense grid of sub-pixels that can be electronically controlled to mask or unmask the incident light. As such, multiplexed encoding of a given scene can be performed by programming the SLM to create time-varying masking patterns \cite{SPI}. The resultant multiplexed measurements can then be used to solve an inverse problem for reconstructing the HR image of the visual scene \cite{fpa-cs}.

Multiplexed imaging holds particular promise for performant imaging in relatively high-frequency bands where manufacturing HR sensors is challenging \cite{SPI_hist, fpa-cs, Mahalanobis2014,stantchev2020thz}. While systems can be implemented via single-pixel sensors (i.e., photo-detectors), recent studies have adopted LR albeit multi-pixel focal plane array (FPA) sensors for their favorable performance/cost trade-off  \cite{fpa-cs, cassi, wagadarikar2008spectral}. Yet, in compressive FPA systems, measurements can be corrupted by system non-idealities including optical aberrations, distortion and vibration, which in turn compromise image quality \cite{fpa-cs, arguello2012spatial, jin2023long, cassi, wagadarikar2008spectral}. These non-idealities are particularly problematic towards higher wavelengths, such as those in thermal imaging \cite{wu2019focal}. 

A fundamental approach to cope with system non-idealities is to conduct offline calibration scans to measure the system matrix that describes the relationships between individual pixels in the HR SLM and the LR FPA. In a common albeit slow calibration approach, measurements are taken while a single-pixel of the SLM is unmasked to propagate the incident scene, and systematically traversed across grid locations \cite{Mahalanobis2014, jin2023long}. For efficiency, measurements can also be taken by using random masking patterns illuminating multiple grid locations simultaneously, and an inverse problem can be solved to recover the system matrix \cite{jin2023long}. These previous methods are not applicable when using fixed coded aperture filters, so they necessitate the use of electronically-switchable (ES) SLMs \cite{cassi, sozak2022}. Furthermore, offline calibration scans incur undesirable computational and calibration time burdens, as system matrix sizes are typically large (e.g., a $256^2 \times 1024^2$ system matrix for a $1024 \times 1024$ SLM and a $256 \times 256$ FPA).

Here, we introduce a novel multiplexed imaging system based on online calibration, CalibFPA, that corrects FPA measurements for influences from system non-idealities. CalibFPA projects the incident scene through a fixed coded aperture and focuses the multiplexed scene onto the LR sensor via a relay lens. Multiple measurements are taken while the coded aperture is locomoted by several sub-pixels to create different SLM patterns. Compared to systems with ES-SLMs, our proposed SPI system is compact due to the fixed coded aperture, and it enables the joint design of the piezo-stage with the coded aperture \cite{sozak2022}. For reconstruction, a deep neural network performs online correction of multiplexed measurements, and a plug-and-play algorithm then recovers the HR image from corrected measurements. We consider the correction of optical aberrations, misalignment and transfer function of the relay lens, which are modeled through a point spread function (PSF) \cite{wu2019focal}. As such, CalibFPA avoids the need for offline calibration and the computational burden introduced by the system matrix during reconstruction. Our contributions are as follows:
\begin{itemize}
    \item We propose a compact, low-cost compressive FPA system that acquires multiplexed measurements via piezo-stage-driven locomotion of a fixed coded aperture.
    \item We introduce a novel deep-learning calibration for online correction of compressive FPA measurements against system non-idealities to improve efficiency in scanning and reconstruction.
    \item We derive the theoretical foundation for CalibFPA that relies on a physics-driven approach based on deconvolution of the relay lens PSF.
\end{itemize}

\section{Related Work}

A powerful imaging framework that avoids expensive HR sensors is single-pixel imaging (SPI) \cite{SPI_hist, duarteSPI}. In SPI, multiplexed encoding of HR spatial information is performed via an SLM and the encoded scene is then imaged using a single photo-detector. The SLM selectively masks or unmasks incident light at each pixel, so the encoded scene is expressed as the multiplication of the original scene with a Bernoulli-type matrix. Use of a single-pixel sensor inherently limits unwanted influences from system non-idealities. Thus, leveraging regularization priors, an inverse problem can be solved to recover the HR image of the scene from raw, multiplexed LR measurements \cite{baraniuk2008simple}. SPI systems typically operate at several kHz to permit real-time imaging. Yet, the use of single-pixel measurements inevitably hampers imaging efficiency.  

To improve imaging efficiency, recent studies have considered using a multi-pixel LR sensor instead of a photo-detector \cite{karOsa, lisens, Mahalanobis2014, jin2023long}. A prominent approach combines HR ES-SLMs for spatial encoding with LR FPAs for sensing \cite{fpa-cs}. An alternative approach uses coded apertures along with dispersive prisms to encode information in the spectral domain \cite{cassi, yang2023mid}. Joint encoding across spatial and spectral dimensions has also been proposed \cite{arguello2012spatial, arguelloCassi}. The presence of multiple pixels in FPA systems renders them more susceptible to system non-idealities. Thus, reliable reconstruction characteristically requires the solution of an inverse problem where system non-idealities are effectively accounted for \cite{tval3, shi2023provable, chen2023prior}. 

Previous studies have commonly proposed offline calibration scans to measure a system matrix that enable compensation of system non-idealities during reconstruction. A conventional offline method is point-scan calibration where ES-SLMs are used to mask/unmask a single HR pixel per measurement. Since separate measurements must be taken for each combination of HR SLM and LR FPA pixels, this approach is experimentally burdening \cite{Mahalanobis2014}. An offline method for accelerated calibration employs compressed sensing (CS) on sparse calibration measurements \cite{jin2023long}. Instead of a single HR pixel, a random subset of HR pixels are masked/unmasked per measurement, and a CS algorithm is then used to recover the entire system matrix. That said, both point-scan and CS methods for offline calibration require the use of an electronically controllable SLM, which is not applicable on fixed coded apertures \cite{sozak2022}. Furthermore, the system matrix capturing the relations between individual pixels in the SLM and the FPA is large and dense \cite{fpa-cs, Mahalanobis2014}. This incurs excessive computational burden during image reconstruction, potentially compromising real-time processing capabilities of the imaging system \cite{fpa-cs, Mahalanobis2014, zhang2021high}. A recent study has proposed to improve computational efficiency via a block-wise reconstruction approach \cite{zhang2021high}. The HR pixels in the SLM are split into blocks spanning the size of LR pixels, and a locally-constrained calibration matrix is then measured. Afterwards, reconstruction is attained by processing measurements in each block separately. While promising results have been reported, the locality constraints on the system matrix reduce the sensitivity of this approach to the global context in HR images. 

Here, we introduce a novel compressive FPA system based on online calibration to cope with system non-idealities without introducing excessive computation load. To our knowledge, CalibFPA embodies the first online calibration method for compressive FPA imaging. Unlike studies based on ES-SLM, CalibFPA utilizes a fixed coded aperture and locomotes it via a piezo-stage for super-resolving the spatial dimensions. Unlike offline calibration methods, CalibFPA avoids the need to acquire and store large system matrices during image reconstruction. Instead, it performs online calibration via a deep network to enable efficient reconstructions on corrected measurements. Unlike block-wise methods with locality constraints, CalibFPA processes the entire set of measurements collectively to improve sensitivity to global information. These unique technical attributes enable CalibFPA to produce high-quality reconstructions of HR scenes efficiently.

\section{Theory}

\subsection{Compressive FPA-based Imaging}

\begin{figure}[t]
    \centering
    \includegraphics[width=1\columnwidth]{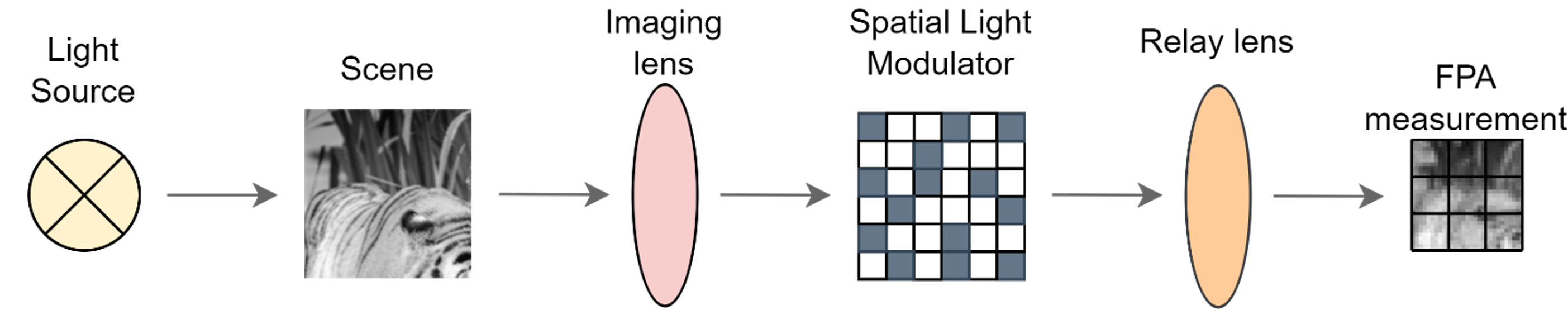}
    \caption{Multiplexed imaging system. An imaging lens focuses incident light coming from the scene onto a high-resolution (HR) spatial light modulator (SLM). A relay lens then focuses the encoded scene onto a low-resolution (LR) focal plane array (FPA) for measurement.}
    \label{fig:systemModel}
    \vspace{-15px}
\end{figure}

\begin{figure*}[t]
    \centering
    \includegraphics[width=0.8\textwidth]{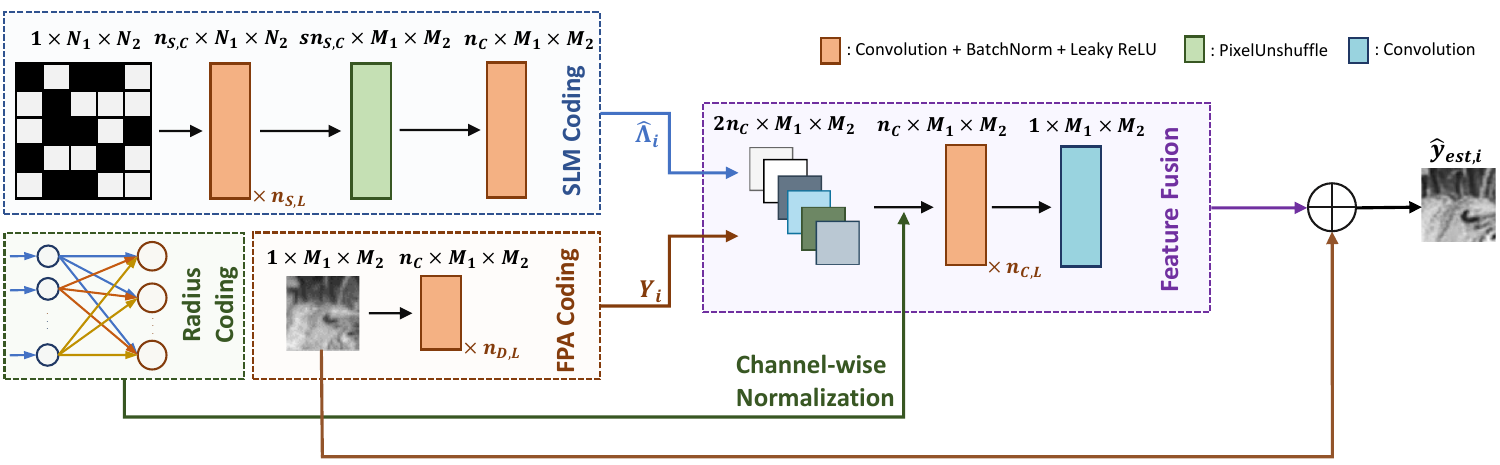}
    \caption{CalibFPA comprises a network architecture with four blocks. The radius coding block receives the Airy disk radius of the relay lens, and computes radius-dependent latent variables via a multi-layer perceptron. The SLM coding block receives the HR SLM pattern and convolutionally encodes it onto LR feature maps. The FPA coding block receives multiplexed LR measurements and convolutionally encodes them onto LR feature maps. The SLM and FPA feature maps are concatenated and weighted according to latent variables from the radius coding block. Afterwards, a feature fusion block equipped with a long-range skip connection from the LR measurements maps the resultant feature map onto corrected measurements.}
    \label{fig:proposed_design}
    \vspace{-10px}
\end{figure*}
In a compressive FPA system, an incident scene is first focused onto a coded aperture for multiplexed encoding of HR information onto LR measurements (Fig.~\ref{fig:systemModel}). The SLM pattern is expressed as a diagonal matrix $\boldsymbol{\Lambda} \in R^{N \times N}$, with entries ``1'' for transparent pixels and ``0'' for opaque pixels. After encoding, a relay lens focuses the scene onto the LR FPA for sensing. Assuming linear shift invariance, the optical effect of the imaging lens can be modeled via a PSF $h_I$ and that for the relay lens via a PSF $h_R$ \cite{dumas2016computational, zhang2021high}:
\begin{align}
    \yb_i = \Db (h_R \circledast (\boldsymbol{\Lambda}_i (h_I \circledast \xb))) + \mathbf{n}_i,
    \label{eq:obsModel}
\end{align}
where $\xb \in R^{N}$ is the HR image vector, $\yb_i \in R^{M}$ are LR measurements, $\mathbf{n}_i \in R^M$ is the noise vector, $\circledast$ is the convolution operator, and $\Db \in R^{M \times N}$ is the box-downsampling matrix. $M$ is the LR sensor size and $N$ is the HR image size, and $i$ denotes the measurements taken with the $i$th coded aperture pattern, $\boldsymbol{\Lambda}_i$. Since only a single coded aperture is printed, multiple patterns are generated by locomoting the coded aperture by several sub-pixels in between each measurement. For a total of $m$ patterns, the resultant super-resolution factor corresponds to $N/M = s$ and the compression ratio is $CR = m/s$ \cite{kar2018performance}. Here, we consider an ideal imaging lens with $h_I = \delta$ as its effects on the system are often relatively moderate \cite{jin2023long}. The relay lens, on the other hand, typically induces significant blur on the encoded HR scene. Its PSF can be described as an Airy disk function with radius $r = 1.22 \lambda f / d$ where $\lambda$ is the wavelength of the light, $f$ is the focal length, $d$ is the diameter of the lens aperture, and $f/d$ corresponds to the F/\# of the relay lens \cite{wu2019focal}. 

Image reconstruction involves solution of an inverse problem based on the forward model. The influences of the lens PSFs and the coded aperture pattern in Eq. \eqref{eq:obsModel} can be condensed in the form of a system matrix as \cite{icip2018, spie_2018}:
\begin{align}
    \yb_i = \mathbf{C}_i \xb + \mathbf{n}_i,
\end{align}
where $\mathbf{C}_i \in R^{M \times N}$ captures the relation between the HR scene and the LR sensor for the $i$th pattern. The regularized inverse problem can then be formulated as:
\begin{align}
    \arg\min_{\xb} R(\xb) \text{ s.t. } \Vert \mathbf{C} \xb - \yb \Vert \leq \epsilon
    \label{eq:inverseProblem}
\end{align}
where $R(\cdot)$ is a regularizer that reflects prior information on the image $\xb$ and $\epsilon$ is a parameter associated with the noise level. Meanwhile, $\mathbf{C} = [\mathbf{C}_1^T \cdots \mathbf{C}_i^T \cdots \mathbf{C}_m^T]^T \in R^{mM \times N}$ denotes the system matrix, and $\mathbf{y} = [\mathbf{y}_1^T \cdots \mathbf{y}_i^T \cdots \mathbf{y}_m^T]^T \in R^{mM}$ denote the measurements. Solution of Eq. \eqref{eq:inverseProblem} typically requires multiple matrix-vector multiplication operations with the system matrix, resulting in $O(mMN)$ complexity.

\subsection{CalibFPA}

\textbf{Online Calibration:}
CalibFPA accounts for undesired influences on measured data $\yb_i$ due to system non-idealities to output corrected data $\hat{\yb}_i$. Since $\Db$ is a box-downsampling operator,  the role of $\Db$ in Eq.~\eqref{eq:obsModel} can be expressed as a strided convolution with box-filter $h_B$ and stride of $s$:
\begin{align}
    \yb_i &= h_B \circledast_{s-strided} (h_R \circledast (\boldsymbol{\Lambda}_i \xb)) + \mathbf{n}_i, \\
     &= \delta \circledast_{s-strided} (h_B \circledast h_R \circledast (\boldsymbol{\Lambda}_i \xb)) + \mathbf{n}_i,
\end{align}
where $\delta$ is the two-dimensional (2D) Dirac-delta function. Correction aims to compensate for the relay lens effects:
\begin{align}
    \hat{\yb}_i &= \delta \circledast_{s-strided} (h_B \circledast (\boldsymbol{\Lambda}_i \xb)). \label{eq:ideal}
\end{align}
Here, we make the key observation that $\hat{\yb}_i$ and $\yb_i$ are related through a convolution, so estimating $\hat{\yb}_i$ from $\yb_i$ does not require knowledge of the HR scene. Thus, a deconvolution based on $h_R$ and $\boldsymbol{\Lambda}_i$ can be adopted to perform the correction.

CalibFPA employs a deep architecture to perform online calibration. The network inputs are LR measurement $\yb_i \in R^{1 \times M_1 \times M_2}$, diagonal elements of the coded aperture pattern reshaped into 2D format $\boldsymbol{\Lambda}_{2D,i} \in R^{1 \times N_1 \times N_2}$ and $r \in R$ value of the Airy disk, with $M_1M_2 = M; N_1N_2 = N$. The network then comprises four blocks for correction.  

\subsubsection{Radius Coding} Since precise estimation of the Airy disk radius with data from an LR sensor is difficult, we pre-defined $9$ radius intervals: $1.5<r<2.5$, $2.5<r<3.5$, $\cdots$, $9.5<r<10.5$. These intervals were one-hot encoded as inputs to a multi-layer perceptron that computes latent variables $\mathbf{r}_i \in R^{2 n_C}$. These latents enable multi-tasking where a single network corrects measurements for all intervals.

\subsubsection{SLM Coding} To encode the HR SLM pattern, a network with $n_{S,L}$ convolutional layers and $n_{S,C}$ channels is used:
\begin{align}
    \mathbf{SLM}_i = c_{s,n_{S,L}}(\cdots(c_{s,1}(\boldsymbol{\Lambda}_{2D,i}))),
\end{align}
where $c_{\cdot, L}(\cdot)$ represents convolution followed by batch normalization and leaky rectified linear unit (ReLU) activation function. 
As $\mathbf{SLM}_i \in R^{n_{S,C} \times N_1 \times N_2}$ is $s_1 \times s_2$ (with $N_1 / M_1 = s_1, N_2 / M_2 = s_2$) times higher in dimensionality compared to $\mathbf{Y}_i$, a pixel unshuffler is used to reduce feature map dimensions while boosting the number of channels:
\begin{align}
    \mathbf{SC}_i = \text{PixelUnshuffler}(\mathbf{SLM}_i),
\end{align}
where $\mathbf{SC}_i \in R^{(s n_{S,C}) \times M_1 \times M_2}$. Finally, the number of channels is reduced to $n_C$ via a convolution layer:
\begin{align}
    \hat{\boldsymbol{\Lambda}}_i = c_{reduce}(\mathbf{SC}_i),
\end{align}
with $\hat{\boldsymbol{\Lambda}}_i \in R^{n_C \times M_1 \times M_2}$.

\subsubsection{FPA Coding} LR measurements are encoded via a network of $n_{D,L}$ convolutional layers, $n_{C}$ channels:
\begin{align}
    \mathbf{Y}_i = c_{D,n_L}(\cdots(c_{D,1}(\yb_i)))
\end{align}
where $\mathbf{Y}_i \in R^{n_C \times M_1 \times M_2}$.

\subsubsection{Feature Fusion} Encoded SLM and FPA feature maps, $\mathbf{Y}_i$ and $\mathbf{SLM}_i$, are first concatenated and then scaled with the latents from the radius coding:
\begin{align}
    \mathbf{DC}_i = \text{Concat}(\mathbf{Y}_i, \mathbf{SLM}_i) / \rb_i,
\end{align}
where $\mathbf{DC}_i \in R^{2n_C \times M_1 \times M_2}$. Resultant feature maps are then projected across $n_{C,L}$ convolutional layers with $n_C$ channels. The final output is computed based on the computed feature maps residually combined with the input measurements with a final convolutional layer without any activation, denoted as $c_F(\cdot)$ that reduces the number of channels to $1$:%
\begin{align}
    \hat{\yb}_{est,i} = \yb_i + c_F( c_{C, n_{C,L}}(\cdots (c_{C, 1}(\mathbf{DC}_i) ) ).
\end{align}
\subsubsection{Training loss} Training is performed via an $\ell_1$-loss:
\begin{align}
\arg\min_\theta \Vert \hat{\yb}_{est,i} - \boldsymbol{y}_{calib,i} \Vert_1,
\label{eq:parametric_DL}
\end{align}
where $\hat{\yb}_{est,i} = g_\theta(\boldsymbol{\Lambda}_i, \boldsymbol{y}_i, r_i)$, $g_\theta$ is the overall network with parameters $\theta$, and $\boldsymbol{y}_{calib,i}$ denotes calibrated reference data measured under ideal optical settings (i.e., $h_I = \delta,h_R = \delta$).

\begin{algorithm}[t]
\caption{PP-FPA: A plug-and-play reconstruction for compressive FPA }\label{alg:admmalg}
\begin{algorithmic}
\State $\textbf{Initialize } \zb_0^{(j)} \text{ and } \db_0^{(j)} \text{ for $j=0,1$, choose } \mu \text{, set } n \gets 0$ 
\While{Stopping criterion is not satisfied}
\State $\xb_{n+1}\gets(\Ib+\mathbf{C}^T\mathbf{C})^{-1}(\mathbf{C}^T(\zb_n^{(0)}+\db_n^{(0)})+\zb_n^{(1)}+\db_n^{(1)})$
\State $\zb_{n+1}^{(0)}\gets \Psi_{\iota_{E(\epsilon,\mathbf{I},\yb)}}(\mathbf{C}\xb_{n+1}-\db_{n}^{(0)})$ 
\State $\zb_{n+1}^{(1)}\gets f_{R}(\xb_{n+1}-\db_{n}^{(1)}; \mu)$ 
\State $\db_{n+1}^{(0)} \gets \db_{n}^{(0)}+\zb_{n+1}^{(0)}-\mathbf{C}\xb_{n+1}$ 
\State $\db_{n+1}^{(1)} \gets \db_{n}^{(1)}+\zb_{n+1}^{(1)}-\xb_{n+1}$ 
\State $n\gets n+1$
\EndWhile
\end{algorithmic}
\end{algorithm}

\textbf{Image Reconstruction:}
Here, we devise a plug-and-play method, PP-FPA, based on alternating direction method of multipliers (ADMM) for image reconstruction \cite{icip2018}. PP-FPA splits Eq.~\eqref{eq:inverseProblem} by decoupling data consistency and regularization steps as described in Alg.~\ref{alg:admmalg}. $\zb^{(0)}_n, \zb^{(1)}_n$ represent the dual variables and $\db^{(0)}_n, \db^{(1)}_n$ are Lagrange multipliers for data consistency and regularization steps at iteration $n$, respectively. $\Psi_{\iota_{E(\epsilon,\mathbf{I},\yb)}}(\cdot)$ is the proximal operator for data consistency that projects the input onto an $\ell_2$-norm ball:
\begin{align}
\Psi_{\iota_{E(\epsilon, \Ib, \yb )}} (\mathbf{s}) &= \left\{
\begin{array}{ll}
\mathbf{s} , & 	\text{if } \Vert \mathbf{s} - \yb \Vert_2 \leq \epsilon \\
\yb + \epsilon \frac{\left( \mathbf{s}-\yb \right) }{\Vert \mathbf{s}-\yb\Vert_2}, & 	\text{if } \Vert \mathbf{s} - \yb \Vert_2 > \epsilon
\end{array}
\right.. \label{eq:data_consistency_update}
\end{align}
Finally, $f_R(\cdot)$ is the Moreau proximal operator associated with the regularization prior $R(\cdot)$. Here, we consider two variants of PP-FPA that differ in $f_R(\cdot)$. The first variant omits regularization altogether to compute the least-squares solution in a single step through fast matrix multiplication operations. The second variant uses a deep denoising prior, and expresses $f_R(\cdot)$ as a forward pass through the denoiser network \cite{kamilovPnP, kamilovPnP2,plugandplayKar}. The pre-trained DnCNN available in PyTorch is utilized as an off-the-shelf denoiser \cite{ppmpi,DnCNNPyTorch}. We pre-compute $(\Ib+\mathbf{C}^T\mathbf{C})^{-1}$ to speed up the reconstruction algorithm. Note that for corrected measurements, $\mathbf{C}$ can be simplified to block-diagonal form with $M$ blocks of size $m \times s$ \cite{spie_2018, icip2018}. Thus, the matrix-vector multiplications have complexity of $O(mN)$ instead of $O(mMN)$ for the regular dense system matrix.

\section{Methods}

\subsection{Optical Setup}
\label{sec:opticalSetup}

\begin{figure}[h]
    \centering
    \begin{minipage}{0.85\columnwidth}
    \centering
    \includegraphics[width=1\columnwidth]{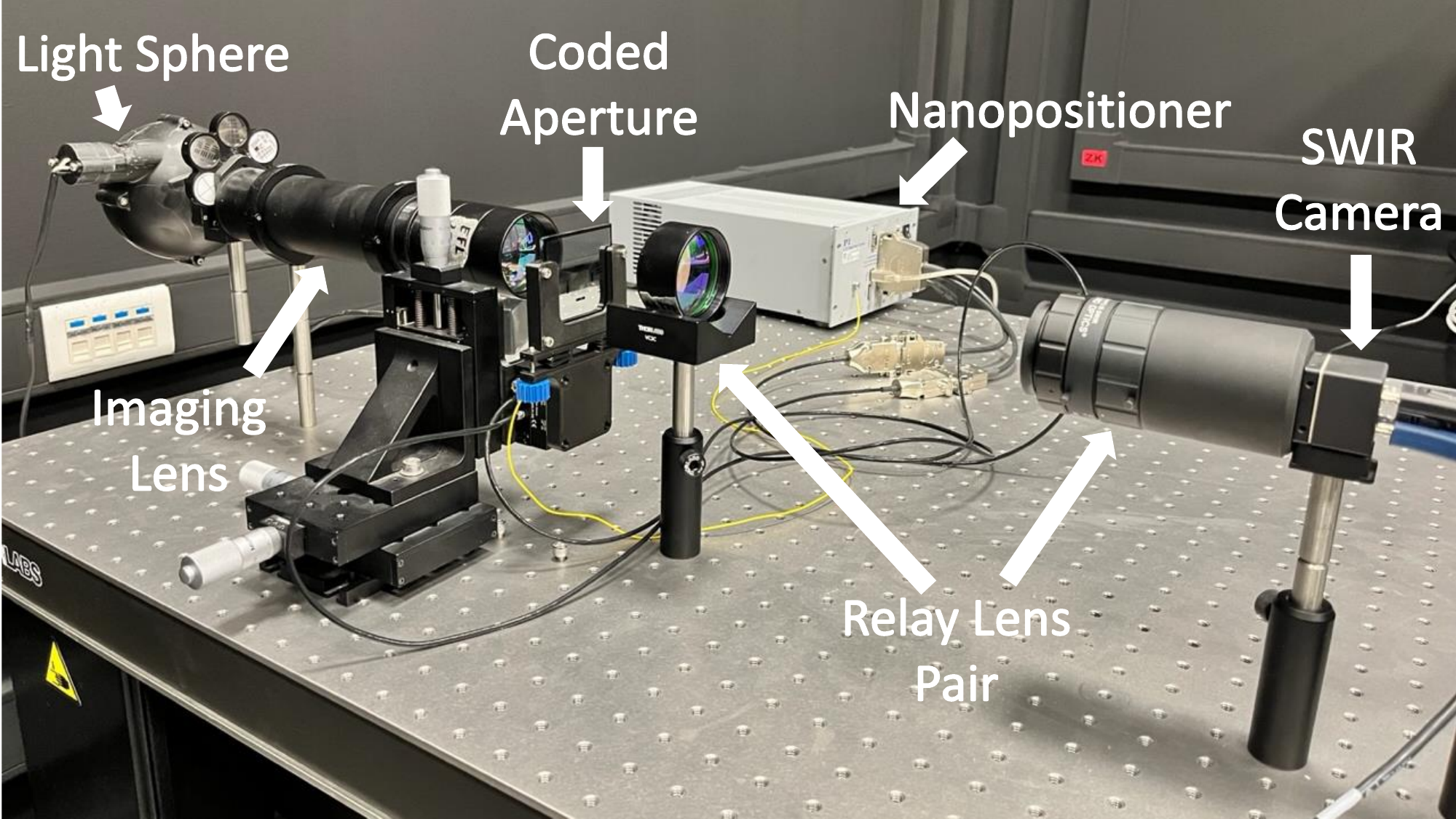}
		\centerline{(a)}
    \vspace{5px}
    \end{minipage}
    \quad \quad \quad
	\begin{minipage}{0.85\columnwidth}
         \centering
		\includegraphics[scale=1.4]{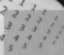}
		\centerline{(b)}
    \end{minipage}
    
    \caption{(a) A photo of the optical setup for the proposed imaging system. A scene illuminated with a SWIR LED is first focused onto a coded aperture via an imaging lens, and then focused onto the FPA sensor via a relay lens. (b) A sample snapshot from the experimental setup.}
\label{fig:systemPhoto}
\end{figure}

To demonstrate CalibFPA, we built the in-house imaging system illustrated in Fig.~\ref{fig:systemPhoto}. A custom SWIR LED light sphere was used for illumination. A Thorlabs AC508-250-C-ML and AC508-100-C-ML lens pair were then used to focus the incident scene onto the coded aperture. The coded aperture was lithographically printed on a soda lime substrate at METU MEMS Center (Ankara, Turkiye). The coded aperture followed a custom pattern of transparent and opaque HR pixels with a pitch of 15 $\mu$m. 
Each 5$\times$5 HR pixel block in the coded aperture corresponded to a single LR pixel on the FPA. Within each block, $20$ transparent and $5$ opaque pixels were selected randomly for an open ratio of $p$=0.8. A closed-loop piezo-stage with PI P622.2CD nanopositioner and PI E-727 Digital controller was used to locomote the coded aperture. In between consecutive measurements, the coded aperture was moved by 15 $\mu$m (i.e., a single HR pixel) in the x-y plane. Following scene encoding, a Thorlabs AC508-250-C-ML and Edmund Optics 50 mm SWIR fixed focal length lens pair was used to relay the encoded scene onto the FPA sensor. The relay lens pair were spaced apart to attain 6x magnification, so the pixel pitch seen on the sensor was 2.5 $\mu$m. An FPA sensor was used comprising Sensors Unlimited 320CSX and a high-frame rate SWIR camera with a resolution of 320$\times$256 and pixel pitch of 12.5 $\mu$m. Hence, the pitch ratio of the camera to the coded aperture resulted in a super-resolution factor of $s$=5$\times$5.

A critical factor for the performance of the imaging system is the alignment between the coded aperture and FPA. We utilized alignment markers given the mismatch between the pixel resolution of the coded aperture and the FPA. Several pixels in the top-left, top-right, bottom-left, and bottom-right parts of the coded aperture were selected for marker placement. The markers were of size 3$\times$3 LR pixels, had a single transparent LR pixel in the middle while remaining pixels were opaque. Alignment was achieved by adjusting the tilt, decenter and position of the objective, the relay lens group and the coded aperture to maximize the ratio of incident light at the central versus peripheral pixels in each marker.

\begin{table}
    \centering
    \caption{Settings for simulation analyses for assessment of the effects of Airy disk radius, input pSNR, $r$ mismatch, SLM mismatch, and offline calibration. SR factor was $s$=5$\times$5 and open ratio was $p$=0.8 in all cases.}
    \label{tab:simulationProc}
\resizebox{0.85\columnwidth}{!}{
    \begin{tabular}{|c|c|c|c|}
        \hline
        Assessment & $\begin{array}{c} \text{Airy Disk} \\ (r) \end{array}$ & $\begin{array}{c} \text{Input pSNR} \\ (dB) \end{array}$ & $\begin{array}{c} \text{Snapshots} \\ (m) \end{array}$  \\
        \hline
        Airy disk & $1.5 < r < 10.5$ & 60 & 5, 25 \\
        \hline
        Input pSNR & $4.5 < r < 8.5$ & 50, 60, 70 & 5 \\
        \hline
        $r$ mismatch & $4.5 < r < 8.5$ & 60 & 5 \\
        \hline
        SLM mismatch & $4.5 < r < 8.5$ & 60 & 5 \\
        \hline
        Offline & $4.5 < r < 8.5$ & 60 & 5 \\
        \hline
    \end{tabular}
}
\end{table}

\subsection{Implementation Details}
To train CalibFPA, a simulated dataset was generated with measurement pairs $\hat{\yb}_i, \yb_i$. Incident scenes were taken as cropped natural images from the DIV2K dataset \cite{div2k}. Randomly generated coded apertures $\boldsymbol{\Lambda}_i$ and a randomly selected $1.5 < r < 10.5$ value for the Airy disk was assumed for each scene. The ratio of transparent to opaque pixels in $\boldsymbol{\Lambda}_i$ was set as 4 to yield an open ratio of $p$=0.8 similar to the coded apertures used in the optical setup. A discrete PSF of size 81$\times$81 was used to implement the Airy disk. A total of $5513, 200, 199$ measurement pairs were simulated for the training, validation, and test sets, respectively. Due to memory considerations, images were cropped to 180$\times$180 for the training/validation sets, albeit to 360$\times$360 for the test set. White Gaussian noise was added at a standard deviation corresponding to a given measurement SNR level. 

Calibration and image reconstruction algorithms were implemented in PyTorch. CalibFPA was trained via the Adam optimizer with a batch size of $64$, an initial learning rate of $10^{-3}$, number of epochs $1000$. The learning rate was scaled by a factor of $0.999$ after each epoch. Hyperparameters were selected to maximize validation performance. Accordingly, $n_{D,L}$=2, $n_{S,L}$=1, $n_{C,L}$=6, $n_{C}$=32, $n_{S,C}$=4 were selected for the architecture. The trained model learned to mitigate not only non-idealities from the Airy disk but also added noise. Hence, $\epsilon$ during image reconstruction was set based on the validation performance of the model. 

\subsection{Competing Methods}
We demonstrated CalibFPA against several online calibration methods as well as offline calibration for reference.

\subsubsection{Raw} Original measurements were not subjected to any correction and directly used in reconstruction. 

\subsubsection{Lucy-Richardson} Original measurements were deconvolved using the LR PSF for the Airy disk  \cite{lucy1974iterative}, and corrected measurements were reconstructed.  

\subsubsection{Blind deconvolution} Deep-learning methods for blind deconvolution were trained to correct measurements on the same datasets separately for each Airy disk radius interval as CalibFPA \cite{blind}. Corrected measurements were reconstructed.

\subsubsection{Offline calibration} The entire system matrix and original measurements were reconstructed with PP-FPA to attain a performance upper bound for image quality. As computing on the entire system matrix was intractable, analyses were restricted to 60$\times$60 test images. Two variants were implemented based on a fully-known PSF (exact radius known), and an inexact PSF for the Airy disk. 





\begin{table}
    \centering
    \caption{Calibration performance of CalibFPA variants and the raw method on the validation set for $s$=5$\times$5. pSNR as mean$\pm$std across the validation set.}
    \label{tab:ablModel}
\resizebox{0.9\columnwidth}{!}{
    \begin{tabular}{|c|c|c|c|c|c|c|}
        \hline
        $r$ & Raw & w/o coding & Strided & CalibFPA-r & \textbf{CalibFPA}  \\
        \hline
         1.5-2.5  & $33.8\pm0.8$& $37.0\pm0.9$& $47.7\pm2.8$& $49.0\pm3.7$& $48.4\pm3.3$\\ 
 \hline 
         2.5-3.5  & $30.9\pm0.9$& $33.8\pm0.9$& $44.9\pm3.2$& $46.3\pm4.1$& $46.0\pm3.9$\\ 
 \hline 
         3.5-4.5  & $29.2\pm0.9$& $32.4\pm0.9$& $43.0\pm3.3$& $44.4\pm4.3$& $44.1\pm4.1$\\ 
 \hline 
         4.5-5.5  & $27.9\pm1.0$& $31.7\pm0.9$& $41.7\pm3.3$& $43.0\pm4.3$& $42.8\pm4.2$\\ 
 \hline 
         5.5-6.5  & $26.9\pm1.0$& $31.1\pm0.9$& $40.4\pm3.3$& $41.6\pm4.2$& $41.5\pm4.1$\\ 
 \hline 
         6.5-7.5  & $26.0\pm1.1$& $30.4\pm0.9$& $39.2\pm3.3$& $40.3\pm4.1$& $40.2\pm4.1$\\ 
 \hline 
         7.5-8.5  & $25.2\pm1.1$& $29.7\pm0.9$& $38.0\pm3.4$& $39.0\pm4.1$& $38.9\pm4.1$\\ 
 \hline 
         8.5-9.5  & $24.6\pm1.1$& $29.0\pm0.9$& $36.9\pm3.4$& $37.7\pm4.1$& $37.6\pm4.0$\\ 
 \hline 
         9.5-10.5 & $24.0\pm1.2$& $28.3\pm0.9$& $35.7\pm3.5$& $36.6\pm4.0$& $36.4\pm4.0$\\ 
 \hline
    \end{tabular}
}
\end{table}


\subsection{Analysis procedures}
Demonstrations were performed on both simulated and experimental data. As the ground-truth images are known for simulated data, quantitative assessments were reported via peak SNR (pSNR) and structural similarity index (SSIM) metrics. Since ground-truth is unavailable in experimental settings, only qualitative assessments were given. 

\subsubsection{Simulations}
Simulated data were generated based on Eq.~\eqref{eq:obsModel} following the same procedures as in the generation of training data. Competing methods were used to correct the measurements. Least-squares reconstructions were performed with $m$=25 snapshots for focused assessment of competing calibration methods. PP-FPA reconstructions were performed to more thoroughly evaluate the overall performance. In the first set of analyses, the effects of the Airy disk radius $r$ 
and input pSNR on the quality of corrected measurements and reconstructed images were examined. In a second set, the influence of mismatch in $r$ and coded aperture pattern were examined. Finally, CalibFPA is compared to offline calibration. Table~\ref{tab:simulationProc} summarizes the settings for the simulation analyses.

\subsubsection{Experiments}
Experimental data were collected with $m$=49 snapshots as described in Sec.~\ref{sec:opticalSetup}. Competing methods were used to correct the measurements, and least-squares and PP-FPA reconstructions were performed. Quality of corrected measurements and reconstructed images were examined for varying radius intervals $r$ and number of snapshots $m$.

\section{Results}

\subsection{Ablation Studies}

We first validated the design elements in CalibFPA through a set of ablation studies on the simulated dataset. To assess the importance of SLM coding, we formed a ``w/o coding'' variant that omitted the coded aperture. To assess the importance of pixel unshuffling in downsampling, we formed a ``Strided'' variant that replaced the unshuffling block with a strided convolution block. To examine the efficacy of multi-tasking across Airy disk radii, we formed a ``CalibFPA-r'' variant that was trained separately at each $r$ value to capture an upper performance bound for CalibFPA. Table~\ref{tab:ablModel} lists performance metrics for variant models at $s$=5$\times$5.

As expected, we find a notable loss in the quality of measurements with the raw method, particularly towards higher Airy disk radii. While w/o SLM Coding improves performance by $3.9$dB compared to raw data on average, CalibFPA significantly outperforms this variant by $10.3$dB. Next, we find that CalibFPA yields $1.0$dB higher pSNR than the ``Strided'' variant. These results suggest that both SLM coding and pixel unshuffling contribute to method performance. Meanwhile, there is only moderate performance difference between CalibFPA and CalibFPA-r, suggesting that CalibFPA is effective in multi-tasking across $r$ values.

\subsection{Analyses on Simulated Data}

We conducted a systematic set of analyses on simulated data to examine the performance of competing methods. 

\subsubsection{Airy Disk Radius}

We evaluated the effect of $r$ on calibration performance. The simulated imaging system assumed $s$=5$\times$5, $p$=0.8, and input pSNR=60 dB. Table~\ref{tab:testRvaluesLR} lists calibration performance for corrected measurements. Calibration performance expectedly drops with increasing $r$ for all methods. Yet, CalibFPA achieves $8.4$dB higher pSNR than the closest competing method on average, and the performance benefits are larger for higher $r$. 

We then evaluated the effect of $r$ on reconstruction performance. Table~\ref{tab:testRvaluesLS} lists performance for $m = 25$ snapshots and least-squares reconstruction. On average, CalibFPA attains an improvement of $8.1$dB pSNR and $31.8$\% SSIM over the top-contending method. Table~\ref{tab:testRvaluesHR} lists performance for $m = 5$ and PP-FPA reconstruction. On average, CalibFPA improves pSNR by $9.6$dB and SSIM by $40.5$\% SSIM over the closest competitor. Note that, for PP-FPA, the performance benefits with CalibFPA-corrected measurements are more noticeable. 

\begin{figure}
    \centering
	\begin{minipage}{0.36\linewidth}
 \centering
		\centerline{Reference Image}
        \vspace{\verticalSpacingDist}
		\includegraphics[scale=\scaleSimulated]{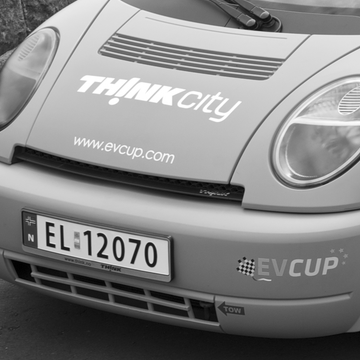}
    \end{minipage}
	\begin{minipage}{0.36\linewidth}
 \centering
		\centerline{Low Res. Image}
        \vspace{\verticalSpacingDist}
		\includegraphics[scale=\scaleSimulated]{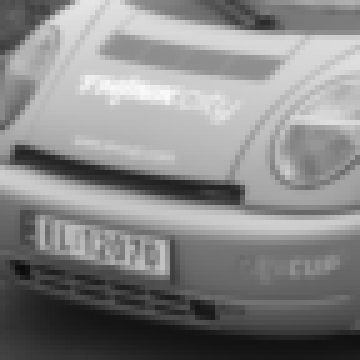}
    \end{minipage}
	\caption{
		A representative image pair depicting the high-resolution reference and low-resolution image from the simulated dataset.}
	\label{fig:simImages}
 \vspace{-10px}
\end{figure}

\begin{figure*}
    \centering

    
    \begin{minipage}{0.01\linewidth}
        \rotatebox{90}{Raw}        
    \end{minipage}
	\begin{minipage}{\MinipageSim\linewidth}
 \centering
		\centerline{$r$: 1.5 - 2.5}
        \vspace{\verticalSpacingDist}
		\includegraphics[scale=\scaleSimulated]{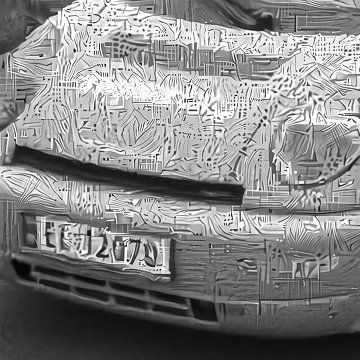}
    \end{minipage}
	\begin{minipage}{\MinipageSim\linewidth}
 \centering
		\centerline{$r$: 3.5 - 4.5}
        \vspace{\verticalSpacingDist}
		\includegraphics[scale=\scaleSimulated]{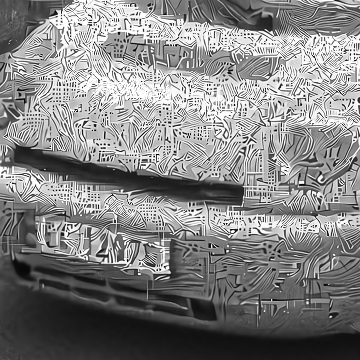}
    \end{minipage}
	\begin{minipage}{\MinipageSim\linewidth}
 \centering
		\centerline{$r$: 5.5 - 6.5}
        \vspace{\verticalSpacingDist}
		\includegraphics[scale=\scaleSimulated]{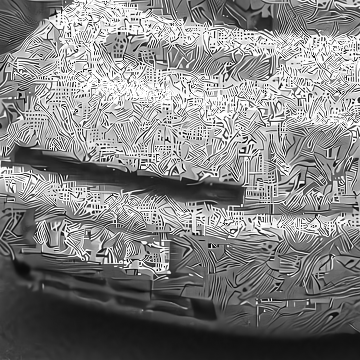}
    \end{minipage}
	\begin{minipage}{\MinipageSim\linewidth}
 \centering
		\centerline{$r$: 7.5 - 8.5}
        \vspace{\verticalSpacingDist}
		\includegraphics[scale=\scaleSimulated]{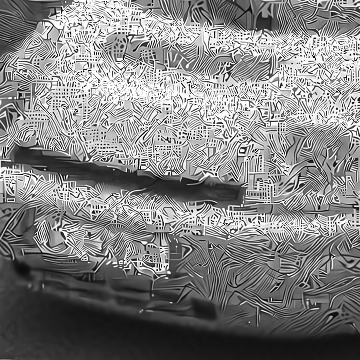}
    \end{minipage}
    \begin{minipage}{\MinipageSim\linewidth}
 \centering
		\centerline{$r$: 9.5 - 10.5}
        \vspace{\verticalSpacingDist}
		\includegraphics[scale=\scaleSimulated]{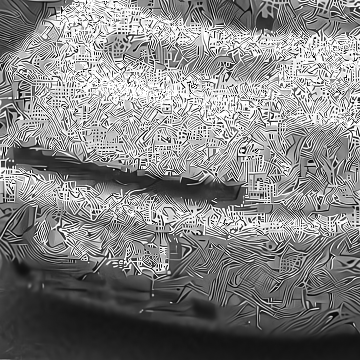}		
    \end{minipage}
    
    \begin{minipage}{0.01\linewidth}
        \rotatebox{90}{Lucy-Richardson}        
    \end{minipage}
	\begin{minipage}{\MinipageSim\linewidth}
 \centering
        \vspace{\verticalSpacingDist}
		\includegraphics[scale=\scaleSimulated]{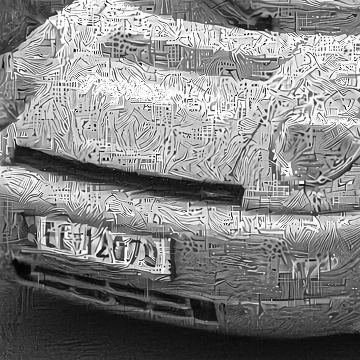}
    \end{minipage}
	\begin{minipage}{\MinipageSim\linewidth}
 \centering
        \vspace{\verticalSpacingDist}
		\includegraphics[scale=\scaleSimulated]{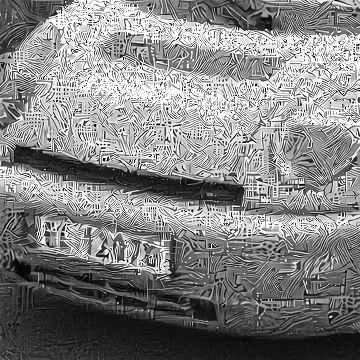}
    \end{minipage}
	\begin{minipage}{\MinipageSim\linewidth}
 \centering
        \vspace{\verticalSpacingDist}
		\includegraphics[scale=\scaleSimulated]{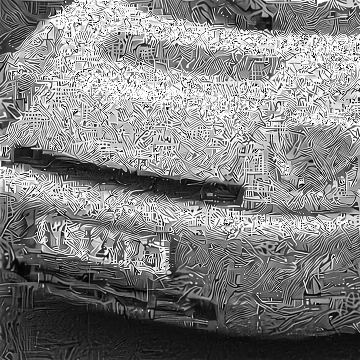}
    \end{minipage}
	\begin{minipage}{\MinipageSim\linewidth}
 \centering
        \vspace{\verticalSpacingDist}
		\includegraphics[scale=\scaleSimulated]{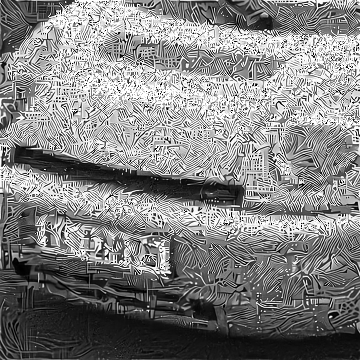}
    \end{minipage}
    \begin{minipage}{\MinipageSim\linewidth}
 \centering
        \vspace{\verticalSpacingDist}
		\includegraphics[scale=\scaleSimulated]{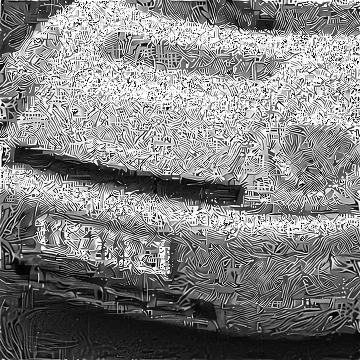}		
    \end{minipage}
    
    \begin{minipage}{0.01\linewidth}
        \rotatebox{90}{Blind}        
    \end{minipage}
	\begin{minipage}{\MinipageSim\linewidth}
 \centering
        \vspace{\verticalSpacingDist}
		\includegraphics[scale=\scaleSimulated]{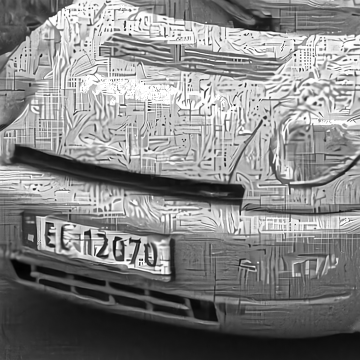}
    \end{minipage}
	\begin{minipage}{\MinipageSim\linewidth}
 \centering
        \vspace{\verticalSpacingDist}
		\includegraphics[scale=\scaleSimulated]{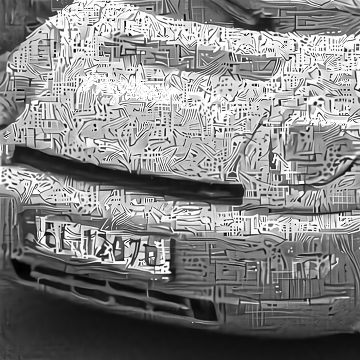}
    \end{minipage}
	\begin{minipage}{\MinipageSim\linewidth}
 \centering
        \vspace{\verticalSpacingDist}
		\includegraphics[scale=\scaleSimulated]{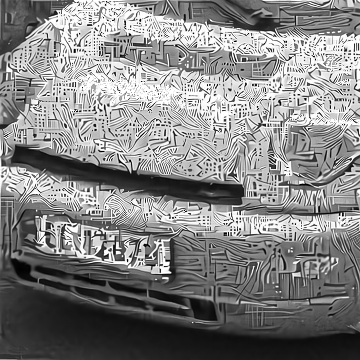}
    \end{minipage}
	\begin{minipage}{\MinipageSim\linewidth}
 \centering
        \vspace{\verticalSpacingDist}
		\includegraphics[scale=\scaleSimulated]{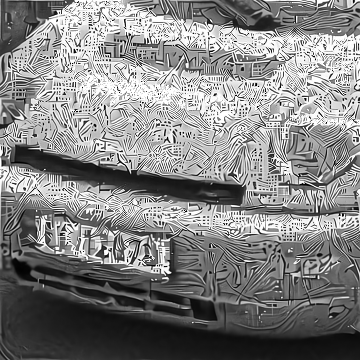}
    \end{minipage}
    \begin{minipage}{\MinipageSim\linewidth}
 \centering
        \vspace{\verticalSpacingDist}
		\includegraphics[scale=\scaleSimulated]{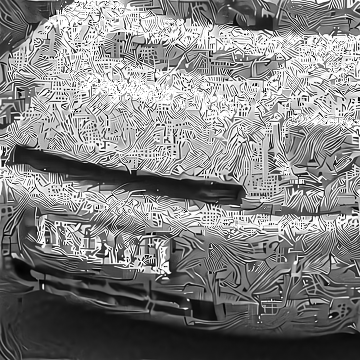}		
    \end{minipage}

    \begin{minipage}{0.01\linewidth}
        \rotatebox{90}{CalibFPA}        
    \end{minipage}
	\begin{minipage}{\MinipageSim\linewidth}
 \centering
        \vspace{\verticalSpacingDist}
		\includegraphics[scale=\scaleSimulated]{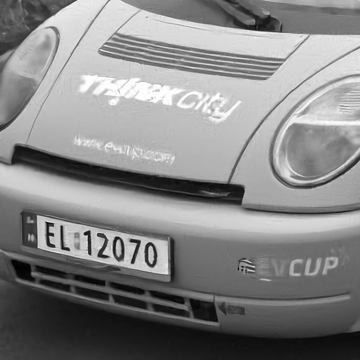}
    \end{minipage}
	\begin{minipage}{\MinipageSim\linewidth}
 \centering
        \vspace{\verticalSpacingDist}
		\includegraphics[scale=\scaleSimulated]{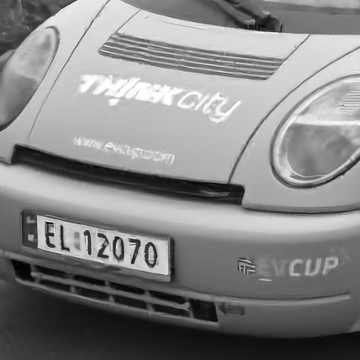}
    \end{minipage}
	\begin{minipage}{\MinipageSim\linewidth}
 \centering
        \vspace{\verticalSpacingDist}
		\includegraphics[scale=\scaleSimulated]{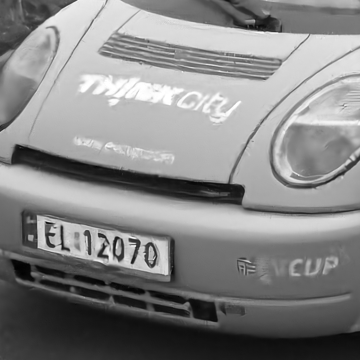}
    \end{minipage}
	\begin{minipage}{\MinipageSim\linewidth}
 \centering
        \vspace{\verticalSpacingDist}
		\includegraphics[scale=\scaleSimulated]{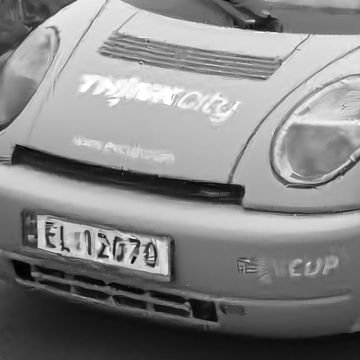}
    \end{minipage}
    \begin{minipage}{\MinipageSim\linewidth}
 \centering
        \vspace{\verticalSpacingDist}
		\includegraphics[scale=\scaleSimulated]{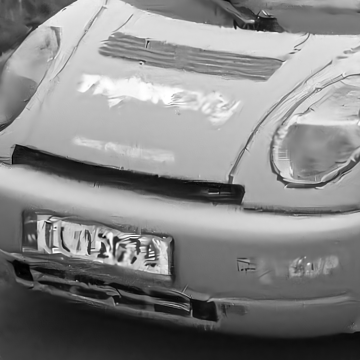}		
    \end{minipage}
	\caption{
		PP-FPA reconstruction of simulated data for $m = 5$ snapshots and varying Airy disk radii and competing methods. }
	\label{fig:simRadiusExp}
 \vspace{-15px}
\end{figure*}

Figure~\ref{fig:simImages} shows the reference image and Fig.~\ref{fig:simRadiusExp} shows image samples reconstructed via PP-FPA for $m = 5$ and representative $r$ values. CalibFPA achieves visible improvements in spatial acuity and noise/artifact suppression against competing methods.

\begin{table}
    \centering
    \caption{Calibration performance of  as a function of Airy disk radius ($r$) for $s$=5$\times$5, $p$=0.8, input pSNR=60dB. pSNR as mean$\pm$std across the test set.}
    \label{tab:testRvaluesLR}
\resizebox{0.75\columnwidth}{!}{
    \begin{tabular}{|c|c|c|c|c|}
        \hline
        $r$ & Raw & Lucy-Rich. & Blind & \textbf{CalibFPA}  \\
        \hline
         1.5-2.5  & $34.9\pm1.1$& $35.2\pm1.1$& $38.6\pm1.2$& $48.0\pm3.5$\\ 
 \hline 
         2.5-3.5  & $31.9\pm1.2$& $32.5\pm1.2$& $35.4\pm1.3$& $45.4\pm4.0$\\ 
 \hline 
         3.5-4.5  & $30.1\pm1.2$& $31.1\pm1.2$& $33.8\pm1.3$& $43.5\pm4.1$\\ 
 \hline 
         4.5-5.5  & $28.8\pm1.3$& $30.3\pm1.2$& $33.1\pm1.3$& $42.2\pm4.0$\\ 
 \hline 
         5.5-6.5  & $27.7\pm1.3$& $29.3\pm1.2$& $32.4\pm1.3$& $40.9\pm3.9$\\ 
 \hline 
         6.5-7.5  & $26.8\pm1.4$& $29.8\pm1.2$& $31.7\pm1.2$& $39.7\pm3.8$\\ 
 \hline 
         7.5-8.5  & $26.0\pm1.4$& $29.5\pm1.2$& $30.9\pm1.3$& $38.4\pm3.8$\\ 
 \hline 
         8.5-9.5  & $25.3\pm1.4$& $28.6\pm1.3$& $30.1\pm1.3$& $37.1\pm3.7$\\ 
 \hline 
         9.5-10.5 & $24.7\pm1.5$& $27.6\pm1.3$& $29.4\pm1.3$& $35.9\pm3.6$\\ 
 \hline 
    \end{tabular}}
\end{table}


\begin{table}
    \centering
    \caption{Least-squares reconstruction performance as a function of Airy disk radius ($r$) for $s$=5$\times$5, $m$=25, $p$=0.8, input pSNR=60dB.}
    \label{tab:testRvaluesLS}
\resizebox{0.9\columnwidth}{!}{
\begin{tabular}{|c|c|c|c|c|c|}
        \hline
        $r$ & Metric & Raw & Lucy-Rich. & Blind & \textbf{CalibFPA}  \\
        \hline
\multicolumn{1}{|c|}{\multirow{2}{*}{1.5-2.5}} & pSNR & $17.5\pm1.1$ & $17.6\pm1.1$ & $19.5\pm1.1$ & $24.7\pm2.5$ \\ \cline{2-6} 
\multicolumn{1}{|c|}{\multirow{2}{*}{}} & SSIM  & $35.8\pm11.6$ & $36.3\pm11.7$ & $45.1\pm12.3$ & $68.6\pm6.2$ \\ 
 \hline 
\multicolumn{1}{|c|}{\multirow{2}{*}{2.5-3.5}} & pSNR & $15.3\pm1.1$ & $15.6\pm1.1$ & $17.1\pm1.1$ & $24.4\pm2.6$ \\ \cline{2-6} 
\multicolumn{1}{|c|}{\multirow{2}{*}{}} & SSIM  & $26.8\pm10.1$ & $28.1\pm10.4$ & $35.4\pm11.6$ & $66.7\pm5.7$ \\ 
 \hline 
\multicolumn{1}{|c|}{\multirow{2}{*}{3.5-4.5}} & pSNR & $14.1\pm1.1$ & $14.6\pm1.1$ & $15.9\pm1.1$ & $24.1\pm2.6$ \\ \cline{2-6} 
\multicolumn{1}{|c|}{\multirow{2}{*}{}} & SSIM  & $21.8\pm8.8$ & $24.1\pm9.4$ & $30.8\pm10.9$ & $64.9\pm5.3$ \\ 
 \hline 
\multicolumn{1}{|c|}{\multirow{2}{*}{4.5-5.5}} & pSNR & $13.4\pm1.1$ & $13.9\pm1.1$ & $15.4\pm1.1$ & $23.8\pm2.6$ \\ \cline{2-6} 
\multicolumn{1}{|c|}{\multirow{2}{*}{}} & SSIM  & $18.4\pm7.8$ & $21.6\pm8.8$ & $28.7\pm10.6$ & $63.1\pm5.0$ \\ 
 \hline 
\multicolumn{1}{|c|}{\multirow{2}{*}{5.5-6.5}} & pSNR & $12.8\pm1.1$ & $13.4\pm1.1$ & $14.9\pm1.1$ & $23.5\pm2.6$ \\ \cline{2-6} 
\multicolumn{1}{|c|}{\multirow{2}{*}{}} & SSIM  & $15.7\pm7.0$ & $19.2\pm8.1$ & $26.2\pm9.8$ & $61.1\pm4.7$ \\ 
 \hline 
\multicolumn{1}{|c|}{\multirow{2}{*}{6.5-7.5}} & pSNR & $12.3\pm1.1$ & $13.5\pm1.1$ & $14.4\pm1.1$ & $23.1\pm2.6$ \\ \cline{2-6} 
\multicolumn{1}{|c|}{\multirow{2}{*}{}} & SSIM  & $13.5\pm6.3$ & $20.2\pm8.3$ & $24.1\pm9.3$ & $58.2\pm4.3$ \\ 
 \hline 
\multicolumn{1}{|c|}{\multirow{2}{*}{7.5-8.5}} & pSNR & $11.8\pm1.1$ & $13.2\pm1.1$ & $13.8\pm1.1$ & $22.6\pm2.6$ \\ \cline{2-6} 
\multicolumn{1}{|c|}{\multirow{2}{*}{}} & SSIM  & $11.8\pm5.7$ & $19.3\pm7.8$ & $21.6\pm8.7$ & $54.6\pm3.9$ \\ 
 \hline 
\multicolumn{1}{|c|}{\multirow{2}{*}{8.5-9.5}} & pSNR & $11.5\pm1.1$ & $12.7\pm1.1$ & $13.3\pm1.1$ & $22.1\pm2.6$ \\ \cline{2-6} 
\multicolumn{1}{|c|}{\multirow{2}{*}{}} & SSIM  & $10.4\pm5.3$ & $17.5\pm7.3$ & $19.3\pm7.8$ & $50.9\pm3.4$ \\ 
 \hline 
\multicolumn{1}{|c|}{\multirow{2}{*}{9.5-10.5}} & pSNR & $11.3\pm1.1$ & $12.2\pm1.1$ & $12.8\pm1.0$ & $21.4\pm2.6$ \\ \cline{2-6} 
\multicolumn{1}{|c|}{\multirow{2}{*}{}} & SSIM  & $9.3\pm5.0$ & $15.5\pm6.7$ & $17.2\pm7.1$ & $46.4\pm3.0$ \\ 
 \hline  
    \end{tabular}
}
\end{table}


\begin{table}
    \centering
    \caption{PP-FPA reconstruction performance as a function of Airy disk radius ($r$) for $s$=5$\times$5, $m$=5, $p$=0.8, input pSNR=60dB.}
    \label{tab:testRvaluesHR}
\resizebox{0.9\columnwidth}{!}{
    \begin{tabular}{|c|c|c|c|c|c|}
        \hline
        $r$ & Metric & Raw & Lucy-Rich. & Blind & \textbf{CalibFPA}  \\
        \hline
\multicolumn{1}{|c|}{\multirow{2}{*}{1.5-2.5 }} & pSNR & $18.3\pm1.6$ & $18.0\pm1.5$ & $19.8\pm1.8$ & $27.8\pm4.7$ \\ \cline{2-6} 
\multicolumn{1}{|c|}{\multirow{2}{*}{}} & SSIM  & $44.9\pm11.8$ & $42.9\pm11.7$ & $51.9\pm11.4$ & $83.7\pm8.3$ \\ 
 \hline 
\multicolumn{1}{|c|}{\multirow{2}{*}{2.5-3.5 }} & pSNR & $16.5\pm1.4$ & $16.3\pm1.3$ & $17.6\pm1.5$ & $27.2\pm4.8$ \\ \cline{2-6} 
\multicolumn{1}{|c|}{\multirow{2}{*}{}} & SSIM  & $36.7\pm12.0$ & $35.6\pm11.9$ & $43.1\pm12.0$ & $82.4\pm9.0$ \\ 
 \hline 
\multicolumn{1}{|c|}{\multirow{2}{*}{3.5-4.5 }} & pSNR & $15.6\pm1.3$ & $15.5\pm1.2$ & $16.6\pm1.4$ & $26.7\pm4.8$ \\ \cline{2-6} 
\multicolumn{1}{|c|}{\multirow{2}{*}{}} & SSIM  & $32.4\pm12.0$ & $32.0\pm11.8$ & $39.2\pm12.1$ & $81.1\pm9.6$ \\ 
 \hline 
\multicolumn{1}{|c|}{\multirow{2}{*}{4.5-5.5 }} & pSNR & $15.0\pm1.3$ & $15.0\pm1.2$ & $16.3\pm1.3$ & $26.4\pm4.7$ \\ \cline{2-6} 
\multicolumn{1}{|c|}{\multirow{2}{*}{}} & SSIM  & $29.6\pm12.0$ & $29.8\pm11.8$ & $37.9\pm12.3$ & $80.0\pm9.9$ \\ 
 \hline 
\multicolumn{1}{|c|}{\multirow{2}{*}{5.5-6.5 }} & pSNR & $14.6\pm1.3$ & $14.5\pm1.2$ & $15.9\pm1.3$ & $26.1\pm4.7$ \\ \cline{2-6} 
\multicolumn{1}{|c|}{\multirow{2}{*}{}} & SSIM  & $27.3\pm11.9$ & $27.5\pm11.6$ & $36.4\pm12.2$ & $78.8\pm10.3$ \\ 
 \hline 
\multicolumn{1}{|c|}{\multirow{2}{*}{6.5-7.5 }} & pSNR & $14.2\pm1.3$ & $14.7\pm1.2$ & $15.6\pm1.3$ & $25.7\pm4.7$ \\ \cline{2-6} 
\multicolumn{1}{|c|}{\multirow{2}{*}{}} & SSIM  & $25.3\pm11.7$ & $29.5\pm11.7$ & $34.8\pm12.3$ & $77.2\pm10.7$ \\ 
 \hline 
\multicolumn{1}{|c|}{\multirow{2}{*}{7.5-8.5 }} & pSNR & $13.9\pm1.3$ & $14.5\pm1.2$ & $15.2\pm1.3$ & $25.1\pm4.6$ \\ \cline{2-6} 
\multicolumn{1}{|c|}{\multirow{2}{*}{}} & SSIM  & $23.7\pm11.6$ & $29.0\pm11.7$ & $32.6\pm12.2$ & $75.0\pm11.3$ \\ 
 \hline 
\multicolumn{1}{|c|}{\multirow{2}{*}{8.5-9.5 }} & pSNR & $13.7\pm1.3$ & $14.1\pm1.2$ & $14.8\pm1.2$ & $24.5\pm4.5$ \\ \cline{2-6} 
\multicolumn{1}{|c|}{\multirow{2}{*}{}} & SSIM  & $22.3\pm11.5$ & $27.2\pm11.6$ & $30.8\pm12.1$ & $72.4\pm12.0$ \\ 
 \hline 
\multicolumn{1}{|c|}{\multirow{2}{*}{9.5-10.5}} & pSNR & $13.5\pm1.3$ & $13.7\pm1.2$ & $14.4\pm1.2$ & $23.6\pm4.4$ \\ \cline{2-6} 
\multicolumn{1}{|c|}{\multirow{2}{*}{}} & SSIM  & $21.2\pm11.3$ & $25.2\pm11.5$ & $28.8\pm12.0$ & $69.3\pm12.7$ \\ 
 \hline  
    \end{tabular}
    }
\end{table}

\subsubsection{Input pSNR}

Next, we evaluated the effect of input pSNR. Table~\ref{tab:snrTestLR} lists calibration performance while Table~\ref{tab:snrTestHR} lists PP-FPA reconstruction performance for competing methods. For each method, there are relatively modest performance differences across input pSNR levels. Yet, CalibFPA yields an average improvement of $8.0$dB in calibration, and $9.4$dB pSNR and $39.9$\% SSIM over the top contending method.

\begin{table}
    \centering
    \caption{Calibration performance as a function of input PSNR for $s$=5$\times$5, $p$=0.8, input pSNR=60dB, $4.5 < r < 8.5$.}
    \label{tab:snrTestLR}
    \resizebox{0.725\columnwidth}{!}{
    \begin{tabular}{|c|c|c|c|c|}
        \hline
        pSNR & Raw & Lucy-Rich. & Blind & \textbf{CalibFPA}  \\
        \hline
         50 & $27.3\pm1.7$& $29.6\pm1.3$& $31.7\pm1.5$& $38.8\pm3.4$\\ 
 \hline 
         60 & $27.3\pm1.7$& $29.7\pm1.3$& $32.1\pm1.5$& $40.3\pm4.1$\\ 
 \hline 
         70 & $27.3\pm1.7$& $29.8\pm1.3$& $32.1\pm1.5$& $40.6\pm4.4$\\ 
         \hline
    \end{tabular}}
\end{table}


\begin{table}
    \centering
    \caption{PP-FPA reconstruction performance as a function of input PSNR for $s$=5$\times$5, $p$=0.8, input pSNR=60dB, $4.5 < r < 8.5$.}
    \label{tab:snrTestHR}
\resizebox{0.9\columnwidth}{!}{
    \begin{tabular}{|c|c|c|c|c|c|}
        \hline
        pSNR & Metric & Raw & Lucy-Rich. & Blind & \textbf{CalibFPA}  \\
        \hline
\multicolumn{1}{|c|}{\multirow{2}{*}{50}} & pSNR & $13.7\pm1.2$ & $14.5\pm1.2$ & $15.3\pm1.3$ & $23.1\pm3.9$ \\ \cline{2-6} 
\multicolumn{1}{|c|}{\multirow{2}{*}{}} & SSIM  & $22.1\pm10.8$ & $27.3\pm11.2$ & $32.1\pm11.9$ & $66.5\pm11.0$ \\ 
 \hline 
\multicolumn{1}{|c|}{\multirow{2}{*}{60}} & pSNR & $14.4\pm1.3$ & $14.7\pm1.2$ & $15.7\pm1.4$ & $25.8\pm4.7$ \\ \cline{2-6} 
\multicolumn{1}{|c|}{\multirow{2}{*}{}} & SSIM  & $26.5\pm12.0$ & $28.9\pm11.7$ & $35.4\pm12.4$ & $77.8\pm10.7$ \\ 
 \hline 
\multicolumn{1}{|c|}{\multirow{2}{*}{70}} & pSNR & $13.8\pm1.2$ & $14.7\pm1.2$ & $15.8\pm1.4$ & $26.2\pm4.7$ \\ \cline{2-6} 
\multicolumn{1}{|c|}{\multirow{2}{*}{}} & SSIM  & $22.5\pm11.1$ & $29.1\pm11.8$ & $35.9\pm12.4$ & $78.7\pm10.5$ \\ 
         \hline
    \end{tabular}
    }
\end{table}


\subsubsection{$r$ Mismatch} 
Inaccurate estimation of the Airy disk radius might influence the system performance, so we evaluated the effect of mismatch between assumed and actual $r$ values. Table~\ref{tab:mismatchRLR} lists calibration performance, whereas Table~\ref{tab:mismatchRHR} lists PP-FPA reconstruction performance. On average, we find that a mismatch equivalent to $1$ HR pixel leads to a loss of $2.4$dB in calibration, and losses of $2.3$dB pSNR, $10.1$\% SSIM in reconstruction. Hence, identifying the correct radius interval can be critical to attain optimal performance.

\begin{table}
    \centering
    \caption{Calibration performance as a function of actual (rows) and assumed (columns) $r$ values for $s$=5$\times$5, $p$=0.8, input pSNR=60dB.}
    \label{tab:mismatchRLR}
    \resizebox{0.75\columnwidth}{!}{
    \begin{tabular}{|c|c|c|c|c|}
        \hline
        $r$ & 4.5-5.5 & 5.5-6.5 & 6.5-7.5 & 7.5-8.5  \\
        \hline
         4.5-5.5 & $42.2\pm4.0$& $39.0\pm3.0$& $33.0\pm1.9$& $28.2\pm1.4$\\ 
 \hline 
         5.5-6.5 & $38.7\pm3.3$& $41.0\pm3.9$& $38.1\pm3.0$& $32.2\pm1.8$\\ 
 \hline 
         6.5-7.5 & $34.3\pm2.3$& $37.8\pm3.3$& $39.7\pm3.9$& $37.2\pm3.1$\\ 
 \hline 
         7.5-8.5 & $31.3\pm2.0$& $33.6\pm2.4$& $36.7\pm3.3$& $38.4\pm3.8$\\ 
 \hline 
    \end{tabular}}
\end{table}


\begin{table}
    \centering
    \caption{PP-FPA reconstruction performance as a function of actual (rows) and assumed (columns) $r$ values for $s$=5$\times$5, $m$=5, $p$=0.8, input pSNR=60dB.}
    \label{tab:mismatchRHR}
\resizebox{0.9\columnwidth}{!}{
    \begin{tabular}{|c|c|c|c|c|c|}
        \hline
        $r$ & Metric & 4.5-5.5 & 5.5-6.5 & 6.5-7.5 & 7.5-8.5  \\
        \hline
\multicolumn{1}{|c|}{\multirow{2}{*}{4.5-5.5}} & pSNR & $26.4\pm4.7$ & $23.7\pm3.8$ & $17.2\pm1.7$ & $13.8\pm1.1$ \\ \cline{2-6} 
\multicolumn{1}{|c|}{\multirow{2}{*}{}} & SSIM  & $80.0\pm9.9$ & $69.6\pm7.5$ & $40.5\pm10.2$ & $26.4\pm10.2$ \\ 
 \hline 
\multicolumn{1}{|c|}{\multirow{2}{*}{5.5-6.5}} & pSNR & $24.4\pm3.9$ & $26.1\pm4.7$ & $23.2\pm3.7$ & $16.5\pm1.6$ \\ \cline{2-6} 
\multicolumn{1}{|c|}{\multirow{2}{*}{}} & SSIM  & $69.3\pm8.4$ & $78.8\pm10.3$ & $67.7\pm7.8$ & $37.5\pm10.2$ \\ 
 \hline 
\multicolumn{1}{|c|}{\multirow{2}{*}{6.5-7.5}} & pSNR & $19.4\pm2.0$ & $24.1\pm3.9$ & $25.7\pm4.7$ & $22.6\pm3.7$ \\ \cline{2-6} 
\multicolumn{1}{|c|}{\multirow{2}{*}{}} & SSIM  & $44.7\pm9.7$ & $68.5\pm9.1$ & $77.2\pm10.7$ & $66.1\pm8.5$ \\ 
 \hline 
\multicolumn{1}{|c|}{\multirow{2}{*}{7.5-8.5}} & pSNR & $16.7\pm1.4$ & $19.0\pm2.0$ & $23.3\pm3.7$ & $25.1\pm4.6$ \\ \cline{2-6} 
\multicolumn{1}{|c|}{\multirow{2}{*}{}} & SSIM  & $32.4\pm10.0$ & $43.2\pm9.8$ & $64.8\pm9.1$ & $75.1\pm11.3$ \\ 
 \hline 
    \end{tabular}
}
\end{table}

\subsubsection{SLM Mismatch}
Despite the use of alignment markers, there can remain small shifts between the assumed and actual position of the coded aperture. Thus, we evaluated the effect of pixel shifts between the coded aperture and the FPA. Shifts within $0.5$ and $1$ pixels were separately considered. Table~\ref{tab:mismatchShiftLR} lists calibration performance while Table~\ref{tab:mismatchShiftHR} lists PP-FPA reconstruction performance for CalibFPA as well as variants retrained on the shifted data. While CalibFPA incurs moderate performance loss under shifted apertures, it performs competitively with retrained variants.

\begin{table}
    \centering
    \caption{Calibration performance as a function of misalignment due to random SLM shifts for $s$=5$\times$5, $p$=0.8, input pSNR=60dB. CalibFPA-0.5 and CalibFPA-1.0 were retrained on data with respective amounts of random shifts.}
    \label{tab:mismatchShiftLR}
    \resizebox{0.65\columnwidth}{!}{
    \begin{tabular}{|c|c|c|c|}
        \hline
        Shift & CalibFPA & CalibFPA-0.5 & CalibFPA-1.0 \\
        \hline
         No Shift & $40.3\pm4.1$& $39.8\pm3.9$& $39.5\pm3.8$\\ 
 \hline 
         -0.5/0.5 & $39.7\pm3.7$& $39.5\pm3.7$& $39.3\pm3.7$\\ 
 \hline 
         -1.0/1.0 & $38.4\pm3.0$& $38.5\pm3.2$& $38.6\pm3.3$\\ 
 \hline 
    \end{tabular}}
\end{table}


\begin{table}
    \centering
    \caption{PP-FPA reconstruction performance as a function of misalignment due to random SLM shifts for $s$=5$\times$5, $p$=0.8, input pSNR=60dB.}
    \label{tab:mismatchShiftHR}
    \resizebox{0.775\columnwidth}{!}{
    \begin{tabular}{|c|c|c|c|c|}
        \hline
        Shift & Metric & CalibFPA & CalibFPA-0.5 & CalibFPA-1.0 \\
        \hline
\multicolumn{1}{|c|}{\multirow{2}{*}{No Shift}} & pSNR & $25.8\pm4.7$ & $25.4\pm4.5$ & $25.0\pm4.3$ \\ \cline{2-5} 
\multicolumn{1}{|c|}{\multirow{2}{*}{}} & SSIM  & $77.8\pm10.7$ & $76.8\pm10.7$ & $76.2\pm10.7$ \\ 
 \hline 
\multicolumn{1}{|c|}{\multirow{2}{*}{-0.5/0.5}} & pSNR & $24.9\pm4.1$ & $24.9\pm4.2$ & $24.8\pm4.2$ \\ \cline{2-5} 
\multicolumn{1}{|c|}{\multirow{2}{*}{}} & SSIM  & $74.0\pm10.1$ & $74.0\pm10.5$ & $74.1\pm10.8$ \\ 
 \hline 
\multicolumn{1}{|c|}{\multirow{2}{*}{-1.0/1.0}} & pSNR & $22.5\pm3.0$ & $23.0\pm3.4$ & $23.3\pm3.6$ \\ \cline{2-5} 
\multicolumn{1}{|c|}{\multirow{2}{*}{}} & SSIM  & $61.5\pm9.3$ & $64.1\pm9.0$ & $66.7\pm10.0$ \\ 
 \hline 
 \end{tabular}}
\end{table}


\subsubsection{Comparison to Offline Calibration}

Finally, we compared the two-stage reconstruction procedure in CalibFPA to end-to-end reconstruction based on offline calibration with the system matrix. Note that these analyses could only be executed on HR images of size $60 \times 60$ to ensure that the offline calibration methods fit in GPU memory. CalibFPA was compared against an ``Ideal'' method that assumed a perfect system with $r \rightarrow 0$, against an exact matrix that assumed perfect knowledge of $r$, and against an inexact matrix that assumed with small perturbations in $r$. Table \ref{tab:fullHR} lists PP-FPA reconstruction performance across different numbers of snapshots. In general, performance for all methods increases with higher $m$. The improvements become marginal for methods with inexact knowledge of $r$ beyond $m$=10. Note that CalibFPA performs comparably to the inexact offline method, while enabling significantly higher computational efficiency.

   

\begin{table}
    \centering
    \caption{PP-FPA reconstruction performance as a function of number of snapshots for $s$=5$\times$5, $p$=0.8, input pSNR=60dB, $4.5 < r < 8.5$.}
    \label{tab:fullHR}
\resizebox{0.9\columnwidth}{!}{
    \begin{tabular}{|c|c|c|c|c|c|}
        \hline
        m & Metric & Ideal & Exact Matrix &  Inexact Matrix & CalibFPA \\
 \hline 
\multicolumn{1}{|c|}{\multirow{2}{*}{1}} & pSNR & $20.7\pm3.1$ & $21.0\pm3.2$ & $20.9\pm3.2$ & $20.8\pm3.1$ \\ \cline{2-6} 
\multicolumn{1}{|c|}{\multirow{2}{*}{}} & SSIM  & $59.1\pm12.8$ & $59.4\pm13.0$ & $59.4\pm12.6$ & $58.3\pm12.8$ \\ 
 \hline 
\multicolumn{1}{|c|}{\multirow{2}{*}{5}} & pSNR & $25.9\pm3.6$ & $24.2\pm3.1$ & $22.1\pm2.4$ & $22.2\pm3.2$ \\ \cline{2-6} 
\multicolumn{1}{|c|}{\multirow{2}{*}{}} & SSIM  & $85.2\pm6.6$ & $78.9\pm7.0$ & $70.8\pm8.2$ & $71.2\pm9.1$ \\ 
 \hline 
\multicolumn{1}{|c|}{\multirow{2}{*}{10}} & pSNR & $29.6\pm3.2$ & $26.7\pm2.5$ & $23.4\pm1.9$ & $23.1\pm3.0$ \\ \cline{2-6} 
\multicolumn{1}{|c|}{\multirow{2}{*}{}} & SSIM  & $92.8\pm3.3$ & $85.2\pm5.0$ & $74.7\pm9.8$ & $76.3\pm6.8$ \\ 
 \hline 
\multicolumn{1}{|c|}{\multirow{2}{*}{15}} & pSNR & $31.9\pm2.6$ & $27.6\pm1.9$ & $23.0\pm1.5$ & $22.9\pm2.9$ \\ \cline{2-6} 
\multicolumn{1}{|c|}{\multirow{2}{*}{}} & SSIM  & $95.5\pm2.0$ & $86.2\pm6.6$ & $73.0\pm11.8$ & $76.5\pm6.1$ \\ 
 \hline 
    \end{tabular}
    }
\end{table}

\subsection{Analyses on Experimental Data}
\begin{figure*}
    \centering
    
    \begin{minipage}{0.01\linewidth}
        \rotatebox{90}{Raw}        
    \end{minipage}
	\begin{minipage}{\MinipageExp\linewidth}
 \centering
        \centerline{$r$: 1.5 - 2.5}
        \vspace{\verticalSpacingDist}
		\includegraphics[scale=\scaleExperimental]{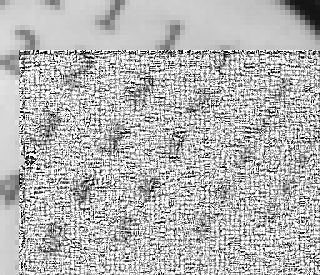}
    \end{minipage}
	\begin{minipage}{\MinipageExp\linewidth}
 \centering
        \centerline{$r$: 4.5 - 5.5}
        \vspace{\verticalSpacingDist}
		\includegraphics[scale=\scaleExperimental]{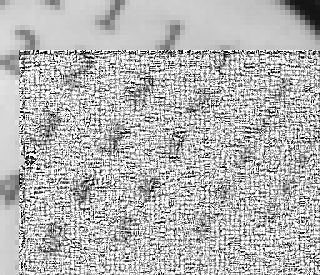}
    \end{minipage}
	\begin{minipage}{\MinipageExp\linewidth}
 \centering
        \centerline{$r$: 5.5 - 6.5}
        \vspace{\verticalSpacingDist}
		\includegraphics[scale=\scaleExperimental]{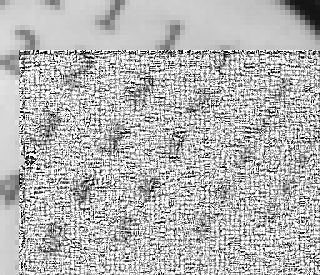}
    \end{minipage}
	\begin{minipage}{\MinipageExp\linewidth}
 \centering
        \centerline{$r$: 6.5 - 7.5}
        \vspace{\verticalSpacingDist}
		\includegraphics[scale=\scaleExperimental]{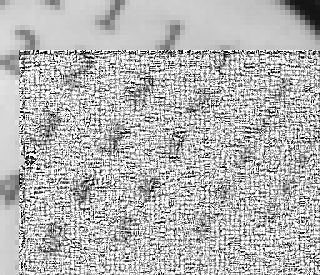}
    \end{minipage}
    \begin{minipage}{\MinipageExp\linewidth}
 \centering
        \centerline{$r$: 9.5 - 10.5}
        \vspace{\verticalSpacingDist}
		\includegraphics[scale=\scaleExperimental]{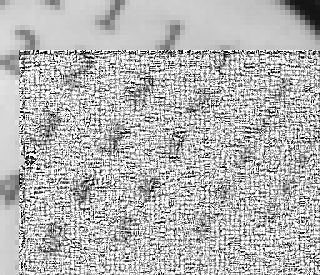}
    \end{minipage}
    
    \begin{minipage}{0.01\linewidth}
        \rotatebox{90}{Lucy-Richardson}        
    \end{minipage}
	\begin{minipage}{\MinipageExp\linewidth}
 \centering
        \vspace{\verticalSpacingDist}
		\includegraphics[scale=\scaleExperimental]{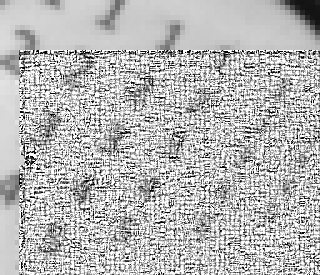}
    \end{minipage}
	\begin{minipage}{\MinipageExp\linewidth}
 \centering
        \vspace{\verticalSpacingDist}
		\includegraphics[scale=\scaleExperimental]{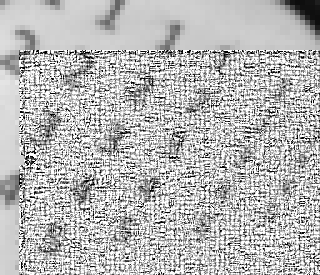}
    \end{minipage}
	\begin{minipage}{\MinipageExp\linewidth}
 \centering
        \vspace{\verticalSpacingDist}
		\includegraphics[scale=\scaleExperimental]{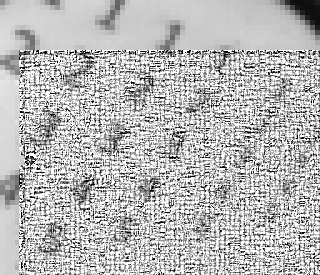}
    \end{minipage}
	\begin{minipage}{\MinipageExp\linewidth}
 \centering
        \vspace{\verticalSpacingDist}
		\includegraphics[scale=\scaleExperimental]{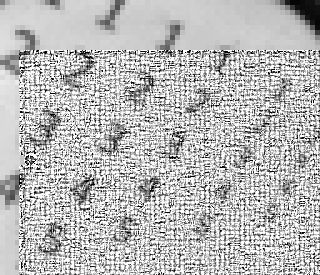}
    \end{minipage}
    \begin{minipage}{\MinipageExp\linewidth}
 \centering
        \vspace{\verticalSpacingDist}
		\includegraphics[scale=\scaleExperimental]{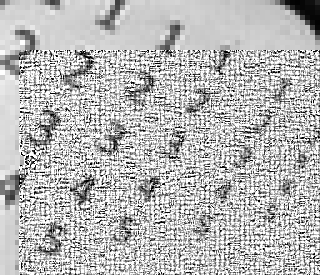}
    \end{minipage}
    
    \begin{minipage}{0.01\linewidth}
        \rotatebox{90}{Blind}        
    \end{minipage}
	\begin{minipage}{\MinipageExp\linewidth}
 \centering
        \vspace{\verticalSpacingDist}
		\includegraphics[scale=\scaleExperimental]{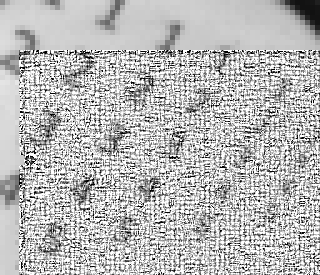}
    \end{minipage}
	\begin{minipage}{\MinipageExp\linewidth}
 \centering
        \vspace{\verticalSpacingDist}
		\includegraphics[scale=\scaleExperimental]{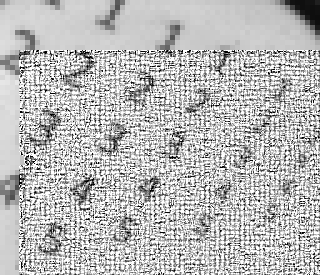}
    \end{minipage}
	\begin{minipage}{\MinipageExp\linewidth}
 \centering
        \vspace{\verticalSpacingDist}
		\includegraphics[scale=\scaleExperimental]{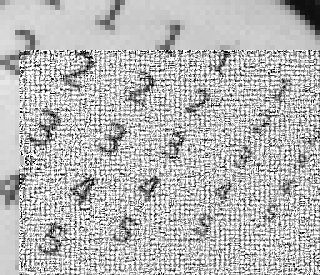}
    \end{minipage}
	\begin{minipage}{\MinipageExp\linewidth}
 \centering
        \vspace{\verticalSpacingDist}
		\includegraphics[scale=\scaleExperimental]{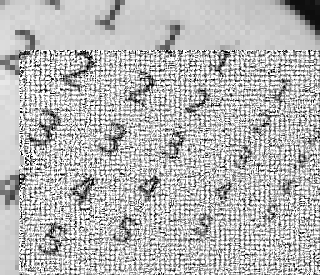}
    \end{minipage}
    \begin{minipage}{\MinipageExp\linewidth}
 \centering
        \vspace{\verticalSpacingDist}
		\includegraphics[scale=\scaleExperimental]{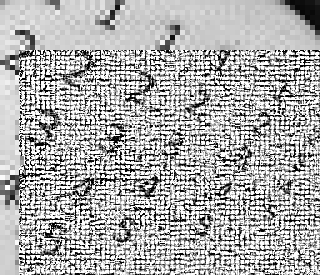}
    \end{minipage}

    \begin{minipage}{0.01\linewidth}
        \rotatebox{90}{CalibFPA}        
    \end{minipage}
	\begin{minipage}{\MinipageExp\linewidth}
 \centering
        \vspace{\verticalSpacingDist}
		\includegraphics[scale=\scaleExperimental]{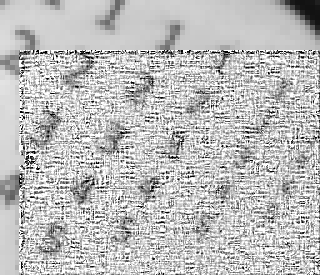}
    \end{minipage}
	\begin{minipage}{\MinipageExp\linewidth}
 \centering
        \vspace{\verticalSpacingDist}
		\includegraphics[scale=\scaleExperimental]{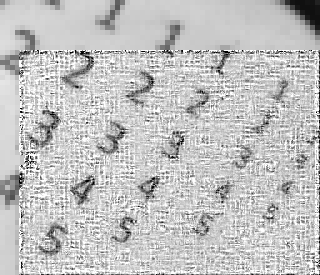}
    \end{minipage}
	\begin{minipage}{\MinipageExp\linewidth}
 \centering
        \vspace{\verticalSpacingDist}
		\includegraphics[scale=\scaleExperimental]{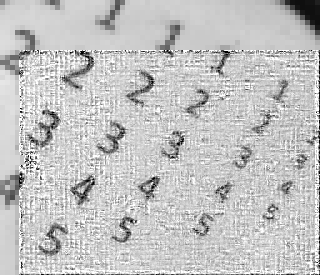}
    \end{minipage}
	\begin{minipage}{\MinipageExp\linewidth}
 \centering
        \vspace{\verticalSpacingDist}
		\includegraphics[scale=\scaleExperimental]{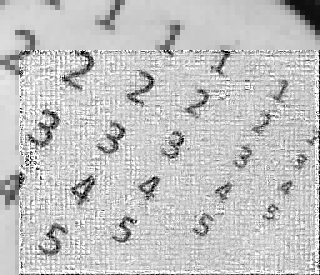}
    \end{minipage}
    \begin{minipage}{\MinipageExp\linewidth}
 \centering
        \vspace{\verticalSpacingDist}
		\includegraphics[scale=\scaleExperimental]{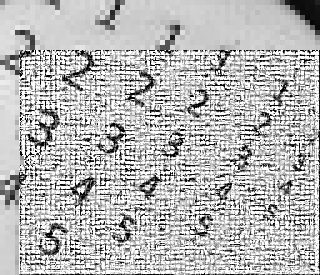}
    \end{minipage}
    
	\caption{
		Least-squares reconstruction of experimental data using competing methods for $m = 49$ snapshots and varying Airy disk radii. }
	\label{fig:experimentalResultLS}
 \vspace{-20px}
\end{figure*}

We also demonstrated CalibFPA on experimental data. Qualitative evaluations were reported as ground truth is unavailable in experimental settings. Fig.~\ref{fig:experimentalResultLS} displays images for varying Airy disk radii based on a least-squares reconstruction and $m = 49$ snapshots. Lucy-Richardson shows a degree of spatial blurring and noise amplification, and blind deconvolution suffers from streak artifacts and noise particularly towards larger radii. In contrast, CalibFPA yields improved image quality in reconstructions with higher spatial acuity in depicting the numerical digits in the scenes, along with improved noise suppression. 

\begin{figure*}
    \centering
    
    \begin{minipage}{0.01\linewidth}
        \rotatebox{90}{Raw}        
    \end{minipage}
	\begin{minipage}{\MinipageExp\linewidth}
 \centering
        \centerline{$m$: 1}
        \vspace{\verticalSpacingDist}
		\includegraphics[scale=\scaleExperimental]{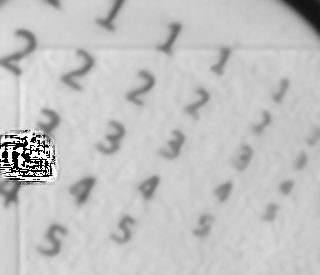}
    \end{minipage}
	\begin{minipage}{\MinipageExp\linewidth}
 \centering
        \centerline{$m$: 5}
        \vspace{\verticalSpacingDist}
		\includegraphics[scale=\scaleExperimental]{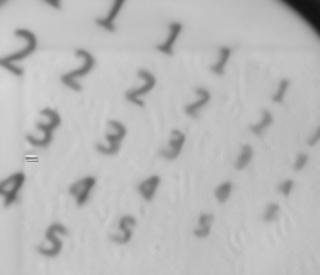}
    \end{minipage}
	\begin{minipage}{\MinipageExp\linewidth}
 \centering
        \centerline{$m$: 10}
        \vspace{\verticalSpacingDist}
		\includegraphics[scale=\scaleExperimental]{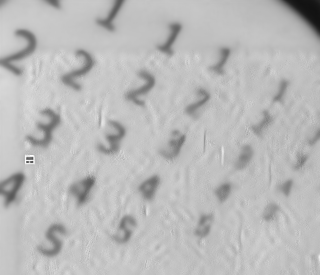}
    \end{minipage}
	\begin{minipage}{\MinipageExp\linewidth}
 \centering
        \centerline{$m$: 15}
        \vspace{\verticalSpacingDist}
		\includegraphics[scale=\scaleExperimental]{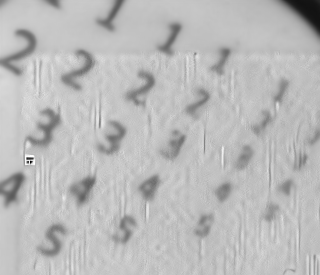}
    \end{minipage}
    
    \begin{minipage}{0.01\linewidth}
        \rotatebox{90}{Lucy-Richardson}        
    \end{minipage}
	\begin{minipage}{\MinipageExp\linewidth}
 \centering
        \vspace{\verticalSpacingDist}
		\includegraphics[scale=\scaleExperimental]{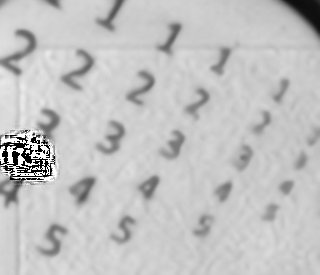}
    \end{minipage}
	\begin{minipage}{\MinipageExp\linewidth}
 \centering
        \vspace{\verticalSpacingDist}
		\includegraphics[scale=\scaleExperimental]{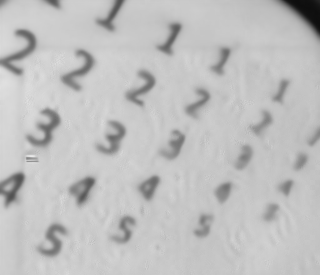}
    \end{minipage}
	\begin{minipage}{\MinipageExp\linewidth}
 \centering
        \vspace{\verticalSpacingDist}
		\includegraphics[scale=\scaleExperimental]{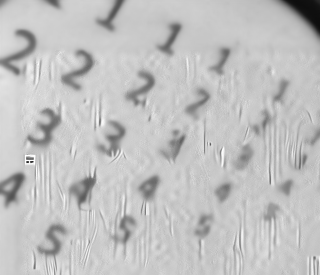}
    \end{minipage}
	\begin{minipage}{\MinipageExp\linewidth}
 \centering
        \vspace{\verticalSpacingDist}
		\includegraphics[scale=\scaleExperimental]{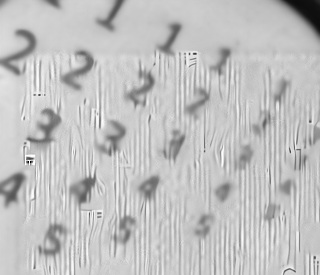}
    \end{minipage}
    
    \begin{minipage}{0.01\linewidth}
        \rotatebox{90}{Blind}        
    \end{minipage}
	\begin{minipage}{\MinipageExp\linewidth}
 \centering
        \vspace{\verticalSpacingDist}
		\includegraphics[scale=\scaleExperimental]{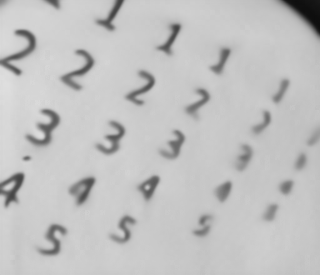}
    \end{minipage}
	\begin{minipage}{\MinipageExp\linewidth}
 \centering
        \vspace{\verticalSpacingDist}
		\includegraphics[scale=\scaleExperimental]{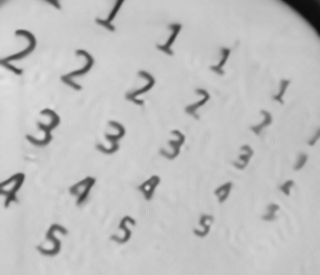}
    \end{minipage}
	\begin{minipage}{\MinipageExp\linewidth}
 \centering
        \vspace{\verticalSpacingDist}
		\includegraphics[scale=\scaleExperimental]{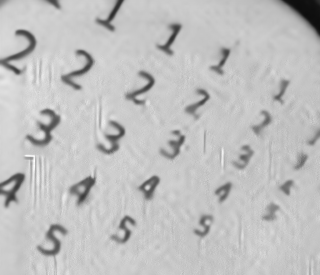}
    \end{minipage}
	\begin{minipage}{\MinipageExp\linewidth}
 \centering
        \vspace{\verticalSpacingDist}
		\includegraphics[scale=\scaleExperimental]{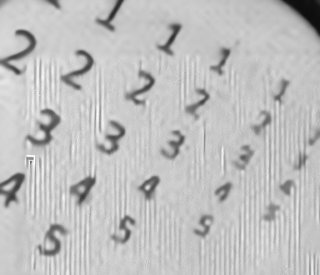}
    \end{minipage}

    \begin{minipage}{0.01\linewidth}
        \rotatebox{90}{CalibFPA}        
    \end{minipage}
	\begin{minipage}{\MinipageExp\linewidth}
 \centering
        \vspace{\verticalSpacingDist}
		\includegraphics[scale=\scaleExperimental]{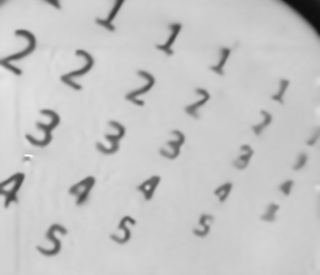}
    \end{minipage}
	\begin{minipage}{\MinipageExp\linewidth}
 \centering
        \vspace{\verticalSpacingDist}
		\includegraphics[scale=\scaleExperimental]{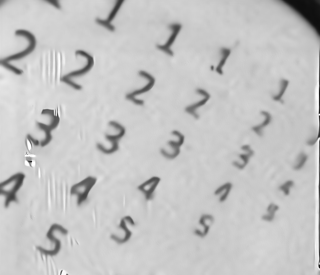}
    \end{minipage}
	\begin{minipage}{\MinipageExp\linewidth}
 \centering
        \vspace{\verticalSpacingDist}
		\includegraphics[scale=\scaleExperimental]{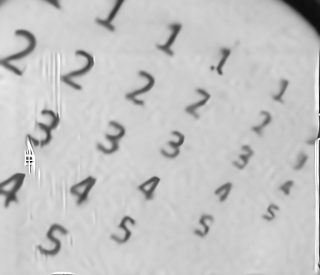}
    \end{minipage}
	\begin{minipage}{\MinipageExp\linewidth}
 \centering
        \vspace{\verticalSpacingDist}
		\includegraphics[scale=\scaleExperimental]{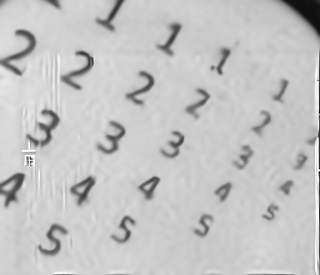}
    \end{minipage}
    
	\caption{
		PP-FPA reconstruction of experimental data using competing methods for $5.5 < r < 6.5$ and varying number of snapshots. }
	\label{fig:experimentalResultPP}
 \vspace{-20px}
\end{figure*}

Next, we examined the performance of competing methods when coupled with the PP-FPA reconstruction for varying numbers of measurement snapshots. Fig.~\ref{fig:experimentalResultPP} depicts results for varying $m$. Reconstructions from all methods show a moderate degree of improvement in spatial acuity for larger $m$, including direct reconstruction of raw data. Yet, competing Lucy-Richardson and blind deconvolution methods yield streaking artifacts and notable noise across the images. Instead, CalibFPA achieves higher quality reconstructions with substantially lower artifacts and noise. 

\section{Discussion}
Here we reported a low-cost imaging system that performs spatial encoding of HR scenes via a coded aperture actuated by a piezo-stage, and captures multiplexed measurements via an LR FPA. This imaging setup allows for a compact physical footprint while still enabling the recovery of HR images of incident scenes. Unlike offline calibration techniques, CalibFPA can mitigate biases from system nonidealities without the need to conduct separate scans to measure the system matrix. To compensate for the optical blurring introduced by the relay lens, CalibFPA uniquely employs an online deep-leaning calibration that performs correction of LR measurements from the FPA. The benefits of online calibration might be particularly evident at higher wavelengths, e.g., in mid-wave and long-wave infrared imaging applications \cite{jin2023long}.  

Reconstruction of images from multiplexed FPA measurements involves solution of an ill-posed inverse problem. To account for system nonidealities, previous techniques employ offline calibration to measure the system matrix, and expressing the inverse problem based on this matrix \cite{Mahalanobis2014}. However, dimensions of the system matrix are often large enough to render the inverse problem computationally intractable \cite{jin2023long, fpa-cs}. Instead, CalibFPA uses online calibration to correct compressive FPA measurements on the fly, and these corrected measurements are reconstructed without the need to process a system matrix. This enables expression of the forward model in block-diagonal form, and an ADMM-based plug-and-play reconstruction then efficiently reconstructs HR images. 

We comprehensively evaluated the performance of CalibFPA against offline and online calibration methods. CalibFPA achieved substantially higher performance than online calibration based on traditional or deep methods, regardless of reconstruction algorithm. Meanwhile, CalibFPA demonstrated a reasonable degree of reliability against mismatches between assumed and actual Airy disk radii and SLM shift values. Finally, CalibFPA performed comparably with computationally-burdening offline calibration, especially when realistic inaccuracies are considered in the system matrix. 

Several technical limitations can be addressed to further enhance the utility of CalibFPA. Here, a plug-and-play reconstruction, PP-FPA, was used based on a pre-trained convolutional denoiser. While PP-FPA facilitates implementation by decoupling the regularization prior to the forward model, performance improvements can be viable when sizable training sets are available. First, transformer architectures can be adopted in the denoiser to elevate sensitivity for long-range context across HR images \cite{ViT,gungor2022transms}. Second, unrolled \cite{MoDL,MongaUnrolling} or deep equilibrium models \cite{deq-mpi,willett} can be used that synergistically integrate the regularization prior with the forward model. Third, representational diversity in reconstructed images might be improved by adopting generative models for inverse problem solutions \cite{ho2020denoising,dar2022adaptive,ozbey2022unsupervised}. Fourth, optimized SLM mask patterns can further improve reconstruction quality \cite{kellman2019physics, van2019coding}. Here, CalibFPA was used to reconstruct single static frames of incident scenes. In applications where the incident scene is a temporal stream, both the calibration and image reconstruction stages of CalibFPA can be modified to jointly process multiple frames via recurrent architectures \cite{videoKar, marcia2008compressive}. Lastly, here we assumed a linear-shift-invariant forward model, yet shift-variant models might allow for broader compensation of system nonidealities for potentially improved fidelity.

\section{Conclusion}
In this study, we introduced a novel compressive FPA system leveraging deep learning for online calibration of measurements against non-idealities and for plug-and-play reconstruction of HR images. CalibFPA employs a physics-driven deep network to correct measurements on the fly against non-idealities arising from the relay lens. Since this avoids the need to utilize a large system matrix, CalibFPA improves computational efficiency for image reconstruction against offline calibration techniques. Comprehensive demonstrations on simulated and experimental data clearly indicate that CalibFPA holds great promise for HR imaging with low-cost compressive FPA sensors.

\ifCLASSOPTIONcaptionsoff
  \newpage
\fi


\bibliographystyle{IEEEtran}
\bibliography{refs}




\end{document}